\documentclass[twocolumn]{aastex631}

\graphicspath{{./}{figures/}}
\usepackage{amsmath}
\usepackage{threeparttable}
\usepackage{graphicx}        % For resizing the table
\usepackage{CJK}
\usepackage{tablefootnote}

\newcommand\sref[1]{\hyperref[#1]{\S~\ref*{#1}}}
\newcommand\fref[1]{\hyperref[#1]{Fig.~\ref*{#1}}}
\newcommand\Eqref[1]{Eq.~(\hyperref[#1]{\ref*{#1}})}
\newcommand\eeqref[1]{Eq.~\hyperref[#1]{\ref*{#1}}}
\newcommand\tref[1]{\hyperref[#1]{Table~\ref*{#1}}}
\newcommand\aref[1]{\hyperref[#1]{Appendix~\ref*{#1}}}

\newcommand{\bpsi}{\boldsymbol{\psi}}
\shorttitle{CR scattering rate from displaced emission}
\shortauthors{Roy et al.}
\begin{document}
%\linenumbers
% Title of the paper, and the short title which is used in the headers.
% Keep the title short and informative.
\title{Not Where You Left Them: Displaced $\gamma$-Rays and X-Rays Reveal the Cosmic Ray Scattering Rate}

% The list of authors, and the short list which is used in the headers.
% If you need two or more lines of authors, add an extra line using \newauthor
\correspondingauthor{Manami Roy}
\email{roy.516@osu.edu}
\author[0000-0001-9567-8807]{Manami Roy}
\affiliation{Center for Cosmology and Astro Particle Physics (CCAPP), The Ohio State University, 191 W. Woodruff Avenue, Columbus, OH 43210, USA}
\affiliation{Department of Astronomy, The Ohio State University, 140 W. 18th Ave., Columbus, OH 43210, USA}
\author[0000-0003-3893-854X]{Mark R. Krumholz}
\affiliation{Research School of Astronomy and Astrophysics, Australian National University, Canberra, ACT 2611, Australia}
\author[0000-0002-2036-2426]{Roland M. Crocker}
\affiliation{Research School of Astronomy and Astrophysics, Australian National University, Canberra, ACT 2611, Australia}
\author[0000-0003-2377-9574]{Todd A. Thompson}
\affiliation{Department of Astronomy, The Ohio State University, 140 W. 18th Ave., Columbus, OH 43210, USA}
\affiliation{Center for Cosmology and Astro Particle Physics (CCAPP), The Ohio State University, 191 W. Woodruff Avenue, Columbus, OH 43210, USA}
\\
% These dates will be filled out by the publisher
%\date{Accepted XXX. Received YYY; in original form ZZZ}

% Prints the current year, for the copyright statements etc. To achieve a fixed year, replace the expression with a number. 
%\pubyear{\the\year{}}

% Don't change these lines

%\label{firstpage}
%\maketitle

\begin{abstract}
Modern X-ray and $\gamma$-ray instruments are revealing a growing class of Galactic non-thermal sources whose emission centroids are measurably offset from the nearest plausible sites of cosmic ray (CR) acceleration. Such ``displaced” sources are seen in keV X-rays and TeV-PeV $\gamma$-rays but not in GeV $\gamma$-rays, have hard spectra, and are not associated with gas clumps, suggesting a leptonic origin. We develop a general framework for understanding displacement, whereby relativistic CR electrons (CRe) injected into the interstellar medium (ISM) with a strongly anisotropic pitch-angle distribution propagate a finite distance from their acceleration site before scattering processes isotropise their directions sufficiently for the emission to become visible. We use CR transport simulations to investigate under what circumstances displacement is likely, finding that it requires an initial pitch angle distribution $\lesssim 45^\circ$ wide, a line of sight broadly edge-on to the magnetic field, and that the source be measured in a waveband where emission is dominated by CRe for which the radiative-loss and pitch-angle–scattering timescales are comparable. For typical Galactic conditions the latter condition is satisfied only for CRe energies %$\geq$
$\gtrsim$
10 TeV, explaining why displaced sources appear at X-ray and TeV but not GeV energies. We further show that, when displacement is detected, it allows a direct inference of the CRe pitch-angle scattering rate.
\end{abstract}



%%%%%%%%%%%%%%%%%%%%%%%%%%%%%%%%%%%%%%%%%%%%%%%%%%

%%%%%%%%%%%%%%%%% BODY OF PAPER %%%%%%%%%%%%%%%%%%

\section{Introduction}
\label{S:intro}

%1. a. Review recent observations of displaced TeV gamma-rays
The 1947 discovery of the $\pi$-meson and its propensity to decay into $\gamma$-rays highlighted the potential of $\gamma$-rays as powerful probes of high-energy, charged, non-thermal cosmic ray (CR) particles \citep{Morrison1958}. Unlike CRs, which are easily deflected by magnetic fields and thus lose information about their origin as they propagate through the magnetised interstellar medium (ISM), $\gamma$-rays travel in straight lines and preserve directional information. This property makes observations of $\gamma$-rays and similar non-thermal emission a valuable tool for identifying and studying astrophysical sites where CRs are accelerated, such as supernova remnants, young and old star clusters, and dwarf galaxies (e.g., \citealt{Helder2012, Song2021, Crocker2022,Pandey2024}). %Since the launch of the Fermi $\gamma$-ray Space Telescope, the number of detected $\gamma$-ray pulsars has grown substantially, enabling, for the first time, detailed population studies of these energetic objects. 
%One of the discoveries allowed by these data is the critical role of $\gamma$-ray emission in the energy output of rotation-powered pulsars, and the efficiency with which millisecond pulsars (MSPs) accelerate cosmic rays within their magnetospheres and subsequently inject them into the surrounding interstellar medium.

However, the connection between non-thermal emission and the sites of CR acceleration is not always straightforward, because in some circumstances the non-thermal emission appears to be displaced from the acceleration site. The first and most prominent example of such displaced emission was the detection of TeV $\gamma$-ray emission from the globular cluster Terzan 5 -- likely a result of inverse Compton scattering off electron-positron pairs accelerated by the cluster's large population of millisecond pulsars (MSPs) -- with an apparent offset of approximately 8 pc from the cluster’s optical centre \citep{Hess2011}. Intriguingly, this feature is not observed at lower $\gamma$-ray energies: Terzan 5 is also detected at GeV energies, but the GeV emission is coincident with the optical cluster position. As TeV capabilities have improved, other such examples have been uncovered. \citet{Shin25a} discovered a similar displacement between the globular cluster UKS 1 and its TeV $\gamma$-ray emission in data from the High-Altitude Water Cherenkov (HAWC) Experiment, while \citet{Cao2025} recently reported a new ultra-high-energy (UHE) $\gamma$-ray source detected by the Large High Altitude Air Shower Observatory (LHAASO), 1LHAASO J1740+0948u, with significant emission above 25 TeV that is displaced by $\approx 0.2^\circ$ from the pulsar PSRJ1740+1000. Further discoveries seem likely as current observatories like HAWC and LHASSO continue to take data, and when the Cherenkov Telescope Array (CTA) comes online a few years from now. At a minimum, there are three additional globular clusters -- Terzan 1, Terzan 2, and NGC 6838 -- that are similar to Terzan 5 and UKS 1 in that they are located relatively close to the Galactic plane and have been detected in the GeV band by \textit{Fermi}-LAT \citep{Song2021}. These are too faint to detect at TeV energies with existing facilities, but they seem likely to be observable by the next generation of instruments, and it is reasonable to anticipate that they may show the same displaced emission as Terzan 5 and UKS 1.

Nor is displaced, non-thermal emission exclusive to $\gamma$-rays. Similar morphologies have been observed in other wavelengths associated with high-velocity young pulsars and MSPs. For example, the LHASSO source detected by \citet{Cao2025} was subsequently found to be connected to PSR J1740+1000 by a faint X-ray tail extending over $\approx 0.2^\circ$ ($\approx 5$ pc at the pulsar distance; \citealt{Gagnon25a}). The radio-quiet $\gamma$-ray pulsar J0357+3205 (\textit{Morla}), located $\sim500$~pc from Earth, similarly exhibits a $9'$ ($\sim1.3$~pc) long X-ray tail detected with \textit{Chandra} and \textit{XMM-Newton} \citep{Luca2011,Marelli2013}, with the peak surface brightness offset by $\sim2'$ from the pulsar position. Similarly, the young radio pulsar PSR J1809--1917 shows a faint, $7'$-long non-thermal collimated structure extending from its X-ray pulsar wind nebula (PWN). This tail extends both opposite to the direction of motion and toward the centroid of the nearby TeV source HESS J1809--193, which is itself offset from the pulsar and may be associated with it \citep{Klingler2020}. \citet{Benbow2021} report three nearby supersonic pulsars -- PSR B0355+54, PSR J0357+3205, and PSR J1740+1000 -- with similar extended X-ray tails; their search for TeV $\gamma$-emission from these tails has thus far yielded only upper limits, but it is reasonable to speculate that these will convert to detections in the near future as more sensitive TeV detectors come online.

%1b. Discuss possible mechanisms to explain displacement, including ours, the alternative discussed in the email thread below, and ideas like bending flux tubes.Say why we think out idea is most promising. \\

%The primary scenario that has been proposed to explain the spatially offset non-thermal emission observed in systems like Terzan 5 is pitch angle scattering of initially anisotropic CRes.  

To explain the spatially offset non-thermal emission observed in systems like Terzan 5, \citet{Krumholz2024} propose a scenario whereby cosmic ray electrons (CRe) are accelerated at the bow shock formed as the source of the relativistic pulsar wind moves rapidly through the interstellar medium, and the pitch angles of the accelerated electrons that escape into the magnetotail behind the source have a pitch angle distribution that is highly biased toward orientations parallel to the magnetic field. \citeauthor{Krumholz2024} suggest that this anisotropy could be due to 
preferential escape of
parallel-oriented electrons from  the acceleration region (a mechanism originally suggested by \citealt{Bykov13a,Bykov17a}), but other mechanisms are also possible. In particular, anisotropic pitch-angle distributions can be imprinted directly during acceleration in magnetic reconnection, where parallel electric fields and Fermi-type processes preferentially energize particles along the magnetic field \citep[e.g.,][]{Comisso2023}, as well as at relativistic shocks, where relativistic aberration confines upstream particles to narrow angular cones in the shock frame \citep[e.g.,][]{Kirk1999}. Anisotropic injection may also occur at quasi-parallel shocks, where shock geometry and field-aligned scattering selectively favor particles with small pitch angles \citep[e.g.,][]{Narita2016}. Regardless of the mechanism, the ultimate effect is the same: near the acceleration site where particles first enter the tail, their emission is beamed away from the observer, but as they propagate downstream, scattering processes gradually reorient a fraction of them toward the line of sight, enabling inverse Compton emission to become detectable at an offset location. This naturally accounts for the observed spatial separation between the cluster and the TeV $\gamma$-ray signal.\footnote{\citet{Krumholz2024} also consider two other scenarios for explaining the displaced TeV emission -- reacceleration of electrons by reconnection in the magnetotail, and transport of an anisotropic population along a twisted magnetotail that curves to align with the line of sight -- and show that they are disfavoured on either theoretical or observational grounds for the case of Terzan 5.}

While this mechanism provides a natural explanation for displaced non-thermal emission, we currently lack a quantitative understanding of the circumstances under which displacement occurs, as well as the connection between the properties of the accelerator and the amount of displacement. Developing this understanding is our primary motivation in this work. The questions we seek to answer are: under what circumstances should we be able to observe displaced non-thermal emission? Why has this phenomenon been observed in TeV $\gamma$-rays (and X-rays), but thus far not in GeV $\gamma$-rays, despite the much larger number of known GeV sources? When displaced emission is seen, what can we infer about the CR pitch-angle scattering rate in the system? As future high-resolution X-ray and TeV $\gamma$-ray telescopes expand the sample of displaced sources, our approach offers a pathway to systematically probe CR transport properties across diverse Galactic environments.  

The layout of the remainder of this paper is as follows. In \autoref{S:methods}, we describe our simulation setup and modelling assumptions, and develop basic intuition for the resulting CR distributions and the key diagnostic metrics used throughout the study. \autoref{S:results} presents our primary findings, while \autoref{S:discussion} offers an interpretation of these results, including their broader implications and directions for future work. Finally, in \autoref{S:conclusion}, we summarize our main findings.

\section{Methods}
\label{S:methods}

Here in \autoref{SS:model_system}, we first introduce the model system that we will explore to gain general insight into displaced emission. We describe our numerical method for solving the equations governing that system in \autoref{SS:numerics}, and define metrics of interest on those solutions in \autoref{SS:metrics}.

%IIA. Review criptic simulations, equations solved, assumptions, dimensionless parameters, etc. — basically, bring people who haven’t read the NatAstro paper up to speed. \\
\subsection{Description of the Simulation}
\label{SS:model_system}

The model system we use to explore displaced emission is the same as the one introduced by \citet{Krumholz2024} in their study of Terzan 5. We consider a scenario in which CRe are injected at a point source located at $z=0$ along a uniform magnetic field aligned with the $z$-axis. The CRe are assumed to be injected continuously at a rate $\dot{N}$ with a momentum distribution given by $dn/dp \propto p^{k_p} \exp(-p/p_{\rm cut})$, valid for $p > p_{\rm min}$. For numerical purposes, we will adopt a lower momentum cutoff of $p_{\rm min} = 1$ GeV$/c$ where such a choice is required, but this has no significant impact on the outcome as long as $p_\mathrm{min}$ is small enough that the observed non-thermal emission is dominated by CRe with larger momenta.

Injection is restricted to a limited range of pitch angles defined by an opening angle $\theta_{\rm inj}$, or equivalently, $\mu > \mu_{\rm inj} = \cos\theta_{\rm inj}$. Although the pitch-angle opening angle $\theta_{\rm inj}$ and spectral index $k_p$ may in reality be correlated -- since in the proposed \citet{Bykov17a} mechanism higher-momentum CRe can escape from the acceleration region over a wider pitch angle range -- there is currently no predictive model for this dependence. Thus, we treat $k_p$ and $\theta_{\rm inj}$ as independent parameters\footnote{{\cite{Krumholz2024} showed that there is no noticeable covariance between 
these parameters
in the posterior probability distribution, suggesting that treating them as independent parameters is sufficient and that introducing an explicit covariance would not significantly affect our results.}}. 

{The ambient magnetic field is assumed to be ``laminar" with no turbulent components\footnote{{\cite{Krumholz2024} discussed the effect of extrinsic turbulence in Section 3 of their Supplementary information and conclude that it is unlikely to be important compared to turbulent fluctuations driven by the streaming instability}}.} It provides a guiding structure for particle motion, with particles travelling along field lines at a velocity $v_z = \mu c$, where $\mu = \cos\theta$ is the cosine of the pitch angle. We neglect gyromotion by averaging over the Larmor orbits and impose a reflecting boundary at $z=0$, mimicking the magnetic topology that %confines particles near 
ensures particles can only escape
the source in certain directions. Such a field configuration is expected to arise naturally when the CR accelerator is travelling supersonically through a background medium -- the situation for all of the candidate displaced emission sources discussed in \autoref{S:intro} -- and sweeps up a magnetotail in its wake. 

Pitch-angle scattering is modelled via a constant diffusion coefficient $K_\mu$, acting to redistribute particles in $\mu$ over time. {In principal, $K_\mu$ can exhibit energy dependence. However, for the purposes of this study, we adopt an energy-independent approximation and leave a more complete treatment to future work.} As particles propagate, they also lose energy. {For the TeV $\gamma$-ray energies at which displacement has been observed, losses are almost certainly dominated by synchrotron or inverse Compton emission over ionization, bremsstrahlung and curvature radiation; in the case of Terzan 5 synchrotron dominates everywhere except in the core of the cluster where the stellar radiation field is most intense \citep{Krumholz2024}\footnote{{Even in the cluster core we can safely neglect $\gamma-\gamma$ pair production, as the optical depth $\tau_{\gamma\gamma} \ll 1$ at TeV energies.}}}. In this case the rate at which electrons lose momentum is
\begin{equation}
\frac{dp}{dt} = -\left(\frac{p}{m_e c}\right)^2 \frac{3}{2}(1 - \mu^2) \frac{m_e c}{t_{c,0}},
\end{equation}
where $t_{c,0} = 2 m_e c / 3 \sigma_{\rm T} U_B$ is the characteristic synchrotron cooling time at $p = m_e c$ for a pitch angle-isotropic population, $\sigma_{\rm T}$ is the Thomson cross-section, and $U_B$ is the magnetic energy density. If inverse Compton losses dominate and those losses occur in the Thomson regime, the result is nearly identical (with $U_B$ replaced by $U_R$, the radiation energy density) except that the pitch angle-dependent term $(1-\mu^2)$ is slightly altered -- for an isotropic radiation field it would be replaced simply by $1$, while for an anisotropic field the term would depend on the alignment between the field line and the radiation field. For simplicity, however, we shall focus on pure synchrotron losses, since the effects of different loss mechanisms matter only at the level of factors of order unity.
%\mrtext{We also note that the optical depth to $\gamma$--$\gamma$ pair production in the cluster radiation field is $\tau_{\gamma\gamma} \ll 1$, implying that the globular cluster is effectively transparent to TeV $\gamma$ rays and that internal absorption can be safely neglected.}

For our model system, the time evolution of the CRe distribution function $f(z, p, \mu)$ is governed by 
\begin{eqnarray}
    \frac{\partial f}{\partial t} & = &
    -\mu c \frac{\partial f}{\partial z} + 
    \frac{\partial}{\partial \mu}\left[(1 - \mu^2) K_\mu \frac{\partial f}{\partial \mu} \right] + {}
    \nonumber \\
    & &
    \frac{m_e c}{t_{c,0}} \frac{\partial}{\partial p} \left[\frac{3}{2}(1 - \mu^2) f \right] + \nonumber \\
    & &
    \dot{N} \frac{dn}{dp} \delta(z) \Theta(\mu - \mu_\mathrm{inj}),
    \label{eq:transport}
\end{eqnarray}
where $\Theta(x)$ is the Heaviside function, representing the injection cut-off in pitch angle. Note that in this equation $f$ is the number of CRs per unit linear momentum, rather than per unit volume in momentum space. We impose a no-flux boundary at $z = 0$, an open boundary as $z \to \infty$, and adopt $f=0$ as an initial condition.

In the limit $t\to\infty$, \citet{Krumholz2024} show that the distribution function $f(z,p,\mu)$ that is the solution to this equation approaches a steady state. These steady-state solutions in general involve complex covariances between $z$, $p$, and $\mu$, and this is the fundamental mechanism that generates displaced emission: in the ultrarelativistic regime where emission from CRe is tightly beamed, an observer will only be able to measure emission from CRe within a very narrow range of $\mu$ oriented such that CRe at this pitch angle are pointed toward the observer at some point during their gyroscopic motion. Thus if the maximum of $f(z,p,\mu)$ for a particular $\mu$ occurs at a position $z$ significantly displaced from $z=0$, then an observer who can view CRe only at that $\mu$ will see the resulting emission as displaced from the injection site.

\subsection{Numerical Solutions}
\label{SS:numerics}

Our next task is to numerically compute the steady-state solution to \autoref{eq:transport} using the CR transport solver code \textsc{criptic} \citep{Krumholz2022}. To do so, we first non-dimensionalize the relevant quantities, allowing a single simulation to generate solutions for arbitrary values of $k_p$ and $p_{\rm cut}$. We define
\[
\tau = K_\mu t,\quad \zeta = \frac{K_\mu}{c}\, z,\quad \dot{\mathcal{N}} = \frac{\dot{N}}{K_\mu},\quad q = \frac{p}{p_{\rm eq}},
\]
where $p_{\rm eq} = K_\mu t_{c,0} m_e c$ is the momentum at which pitch-angle scattering and synchrotron cooling occur on comparable timescales. In these variables, \autoref{eq:transport} becomes
\begin{align}
\frac{\partial f}{\partial \tau} &= -\mu \frac{\partial f}{\partial \zeta} 
+ \frac{\partial}{\partial \mu} \Big[(1-\mu^2) \frac{\partial f}{\partial \mu}\Big] 
+ \frac{\partial}{\partial q} \Big[\frac{3}{2}(1-\mu^2) q^2 f\Big] \nonumber\\
&\quad + \dot{\mathcal{N}}\, \frac{dn}{dq} \, \delta(\zeta)\, \Theta(\mu-\mu_\mathrm{inj}),
\label{eq:fpe_nondim}
\end{align}
where $dn/dq \propto q^{-k_p} \exp(-q/q_\mathrm{cut})$ and $q_\mathrm{cut} = p_\mathrm{cut}/p_\mathrm{eq}$. \autoref{eq:fpe_nondim} is a Fokker--Planck equation (FPE), which has a corresponding It\^o stochastic differential equation \citep{Krumholz2022}
\begin{equation} d\chi_j = A_j(\boldsymbol{\chi}) + \left(\sqrt{K_{ij}(\boldsymbol{\chi})}\, dW_i\right)_j,
\label{eq:sde} 
\end{equation}
where $\boldsymbol{\chi} = (\zeta, q, \mu)$ denotes the coordinates of a single probability packet, $j=1,2,3$ indexes the three dimensions, and $d\mathbf{W}$ is a three-dimensional Wiener process, and $\mathbf{A}$ and $\mathbb{K}$ are the drift vector and diffusion tensor, given by
\[
\mathbf{A} = \left(
\begin{array}{c}
\mu \\ \frac{3}{2} q^2(1-\mu^2) \\ -2\mu
\end{array}
\right),\quad 
\mathbb{K} = \left(
\begin{array}{ccc}
0 & 0 & 0 \\
0 & 0 & 0 \\
0 & 0 &  2 (1-\mu^2)
\end{array}
\right),
\]
with reflecting boundary conditions at $\zeta=0$. An ensemble of packets injected according to an initial momentum distribution $dn/dq$ at a constant rate $\dot{\mathcal{N}}$ reproduces the full solution of the FPE.

\autoref{eq:sde} admits self-similar solutions, allowing a single trajectory $(\zeta(\tau), q(\tau), \mu(\tau))$ to be rescaled for different initial momenta $q'$. Since $\zeta$ and $\mu$ do not depend on $q$ and $A_q \propto q^2$, if a trajectory satisfies $q(0)=1$ and $q(\tau_1)=q_1$, then
\begin{equation}
q'(\tau) = \frac{q_1 q'_1 \tau_1}{q_1 \tau_1 + q'_1 (\tau - \tau_1)(1-q_1)}
\end{equation}
is also a solution (with $q'(\tau_1) = q'_1$) for any $q'_1 < q_1/(1-q_1)$. This allows packets initially at $q=1$ to be mapped to any initial momentum distribution $dn/dq'_0$, where the resulting transformed probability distribution of momenta $q'_1$ at time $\tau_1$ would be 
\[
\frac{dn}{dq'_1} = \left(\frac{q'_0}{q'_1}\right)^2 \frac{dn}{dq'_0}.
\]

This approach significantly reduces computational cost by allowing a library of simulations varying only $\mu_{\rm inj}$, from which solutions for arbitrary parameters can be generated. We use \textsc{criptic} \citep{Krumholz2022} to solve \autoref{eq:sde} for CRe sources injecting packets at $\zeta=0$ with $q=1$ and pitch angles $\mu \in (\mu_0,1)$, considering 16 different values of $\mu_0$ for each source, from $-0.5$ to 1 in steps of 0.1. Details of the numerical setup for the simulations are provided in \citet{Krumholz2024}.

The output of the simulation is a set of  probability packet positions, momenta, and weights (i.e., how many individual CRe a given packet represents) at a series of times; we use the final time ({which is at $\tau\gg 1$ and thus large enough for the distribution to have reached steady state}) for all our results in this paper. To convert the packet properties back to a distribution function, we use a kernel density estimate:
\begin{equation}
f(\zeta,q,\mu) = \dot{\mathcal{N}} \sum_i w_i \, G(\zeta-\zeta_i, h_\zeta) \, G(\mu-\mu_i, h_\mu) \, \phi(q \mid q_i),
\label{eq:fsol}
\end{equation}
where $w_i$, $\zeta_i$, and $q_i$ are the statistical weight, dimensionless position, and dimensionless momentum of packet $i$, $G(x,h) = \exp(-x^2/h^2)/I$ is a Gaussian kernel with bandwidth $h$, $I$ is a normalisation factor, and
\begin{equation}
\phi(q \mid q_i) =
\begin{cases}
\left(q_0/q\right)^2 (dn/dq)_{q=q_0}, & q < q_i/(1-q_i) \\
0, & q > q_i/(1-q_i)
\end{cases}
\label{eq:wgt}
\end{equation}
with $q_0 = (1 + q^{-1} - q_i^{-1})^{-1}$ being the initial momentum a packet would need to reach $q$ at the end of the simulation. For numerical evaluation, we adopt a Gaussian kernel bandwidth $h_{z,\mu} = 0.025$ 
%\mrtext{(We choose $h_{z,\mu}$ such that the Gaussian kernel is well resolved on the $\mu$ and $z$-grid.)} 
and normalize the kernel in $\zeta$ as $I_\zeta = \sqrt{\pi}\,h_z$. For the $\mu$ direction, some care is required near the boundaries at $\mu = \pm 1$, since our Gaussian kernels are truncated at these boundaries. We therefore adopt the normalisation
\begin{equation}
    I_\mu = \frac{\sqrt{\pi}\, h_\mu}{2} \left[ \mathrm{erf}\Big(\frac{1-\mu_i}{h_\mu}\Big) - \mathrm{erf}\Big(\frac{-1-\mu_i}{h_\mu}\Big) \right],
\end{equation}
for a packet with pitch angle $\mu_i$, which approximately accounts for the truncation. Even with this accounting, however, the probability distribution of packet positions near the boundaries depends on the details of our numerical implementation for handling the coordinate singularity at $\mu = 1$, which leads to minor artefacts. To avoid these, we restrict ourselves to considering angles $\mu \in [-0.98,0.98]$.

This procedure allows us to compute the steady-state solution $f(\zeta,q,\mu)$ for any of our chosen injection angles $\mu_\mathrm{inj}$, and for any choice of the parameters $k_p$ and $q_\mathrm{cut}$ appearing in \autoref{eq:wgt}. We can obtain solutions for intermediate values of $\mu_{\mathrm{inj}}$ by linear interpolation between grid points.

%IIB. Review basic results; include some simple plots analogous to those in the NatAstro paper showing the distributions of CRs in pitch angle and position for different momenta. \\

%\autoref{f:CR_dist} presents representative solutions to the above transport equation under various conditions. A key feature that emerges from these results is the distinct behavior of the CR distribution as a function of momentum and pitch angle. At higher CR momenta ($q \gtrsim 1$) and for pitch angles satisfying $\mu > \mu_0 \equiv \cos \theta_0$, the peak of the distribution is noticeably shifted towards positive values of $\zeta$, indicating an anisotropic structure. In contrast, for low-momentum CRs ($q \ll 1$), the distribution becomes significantly broader and more symmetric, with a peak near $\zeta = 0$ and minimal dependence on pitch angle. We compare different spectral indices in the top and middle panels and find minimal dependence on the spectral slope. However, increasing $\mu_{\rm inj}$ (bottom panel) leads to a more symmetric distribution toward lower $\zeta$, thereby suppressing any distinguishable displacement.

%This behavior arises from the fact that the synchrotron cooling time scales inversely with CR momentum ($\propto 1/q$). High-energy CRs, therefore, lose energy before they have time to isotropize in pitch angle, retaining the directional asymmetries present at injection. In contrast, low-energy CRs undergo efficient isotropization before significant cooling occurs, resulting in a more uniform and symmetric distribution.

\subsection{Metrics of Interest}
\label{SS:metrics}
%IIC. Define our metrics of interest, and how we compute them: the position of the peak, the amplitude of the peak relative to the isotropic case, and the ratio of the displacement to the width

We have seen that our simple model system is fully characterised by three dimensionless numbers: the opening angle of injection $\mu_\mathrm{inj}$, the electron powerlaw index at injection $k_p$, and the ratio of the cutoff momentum to the momentum at which the synchrotron loss and pitch angle-scattering times are equal, $q_\mathrm{cut}$. These three numbers, to which we will refer generically via the symbol $\bpsi = (\mu_\mathrm{inj}, k_p, q_\mathrm{cut})$, uniquely determine the shape of the CRe distribution function $f(\zeta, q, \mu)$ as a function of the dimensionless position $\zeta$, momentum $q$, and pitch angle $\mu$. We are now in a position to answer some of the fundamental questions raised in \autoref{S:intro}: for what values of the dimensionless parameters that define our system does $f(\zeta, q, \mu)$ have the properties that would lead us to observe displaced emission? When $f$ has such properties, for what ranges of $q$ and $\mu$ would we expect to see this displacement?

To assist in answering these questions, we now introduce a set of diagnostic metrics derived from the steady-state distribution function $f(\zeta, q, \mu)$, which in practice we always evaluate using \autoref{eq:fsol}. Our diagnostics are as follows:

\begin{itemize}
    \item \textbf{Peak location, $\zeta_{\rm max}$:}  
    This is the position along the magnetic field line $\zeta$ at which the CRe distribution function $f(\zeta,\mu,q)$ reaches its maximum for a given set of problem parameters $\bpsi$ and a particular CRe momentum and pitch angle, ($q$, $\mu$). A non-zero value of $\zeta_{\rm max}$ indicates that the emission produced by CRe of momentum $q$ and seen by an observer at pitch angle $\mu$ relative to the field line will be spatially displaced from the injection site. 

    \item \textbf{Sharpness of the displacement, $S$:} Even if the position $\zeta_\mathrm{max}$ at which the distribution function reaches its maximum is displaced away from the origin, this displacement may not be discernable if the distribution function is not sharply peaked; cases where the distribution function is not peaked will appear as an anisotropic tail of emission, rather than as a displaced peak. To quantify this effect, we find the value $\zeta_{1/2}$ for which $f(\zeta,q,\mu)$ falls to half its maximum value, i.e., we solve the implicit equation $f(\zeta_{1/2}, q,\mu) = (1/2) f(\zeta_\mathrm{max},q,\mu)$; if the equation has more than one root, we select the larger one. We then define the sharpness $S = \zeta_\mathrm{max} / 2 (\zeta_{1/2} - \zeta_\mathrm{max})$; values of $S \approx 1$ indicate that $f(\zeta,q,\mu)$ falls sharply at $\zeta > \zeta_\mathrm{max}$, while values $\ll 1$ indicate that the distribution has a broad peak that falls off only slowly away from its maximum, and thus is likely to be seen as a tail rather than a spot of displaced emission. Clearly this parameter is, like $\zeta_\mathrm{max}$, a function of both the problem parameters $\bpsi$ and the CRe momentum and pitch angle, $(q,\mu)$. In practice, to compute $\zeta_{1/2}$ for a given combination of these parameters, we evaluate $f(\zeta,q,\mu)$ on a fine grid of $\zeta$ values and then construct a linear interpolation function from this grid, then carry out numerical root-finding on this interpolated function. 
    
    \item \textbf{Contrast with on-source emission, $C_\mathrm{on}$:} To produce a distinguishable displaced peak, the distribution needs not only to be displaced ($\zeta_\mathrm{max}>0$) and sharp ($S\gtrsim 1$), CRe density at the peak must be noticeably higher than it is on top of the source. To quantify this requirement, we compute the ratio of the peak value to the value at $\zeta=0$, $C_\mathrm{on} = f(\zeta_{\rm max},q,\mu) / f(0,q,\mu)$. A value significantly greater than 1 indicates that the peak of the emission is significantly brighter than the emission directly above the injection site -- an essential condition for displacement to be observationally distinguishable.

    \item \textbf{Contrast with isotropic injection, $C_{\rm iso}$:} We will see in \autoref{S:results} that all of the metrics we have defined thus far tend to increase with $q$, which is intuitively straightforward to understand: CRe with $q \gg 1$ cool faster than they isotropise, and as a result they retain the highly anisotropic distribution with which they are injected. However, these CRe may not be observable, precisely because they cool so quickly that they never have time to scatter into an observer's line of sight. This issue motivates our last metric, $C_\mathrm{iso}$, which measures the value of the CRe distribution integrated over position $\zeta$ at a particular $(q,\mu)$ combination relative to the value we would expect if the CRs had an isotropic pitch angle distribution at all positions, times, and momenta. We show in \aref{A:iso_derive} that for this case the exact solution is
    \begin{equation}
    F_{\rm iso}(q,\mu) = \frac{\dot{\mathcal{N}}}{2q^2}
    \frac{ \,
    \Gamma\!\left(1+k_p, \frac{q}{q_{\rm cut}}\right)}{
    \Gamma\!\left(1+k_p, \frac{q_{\rm min}}{q_{\rm cut}}\right)},
   \end{equation}
    where $\Gamma(a, x)$ the upper incomplete gamma function. We then define
    \begin{equation}
        C_\mathrm{iso} = \frac{\int_0^\infty f(\zeta,q,\mu)\, d\zeta}{F_\mathrm{iso}(q,\mu)}. 
    \end{equation}
    Values significantly below unity indicate that anisotropy reduces the brightness significantly relative to the isotropic case, in which case there might be displaced emission, but too faint to observe.    
\end{itemize}

Together, these four metrics form the core of our analysis framework. We compute them from the numerically obtained $f(\mu,\zeta,q)$ for each combination of $q$ and $\mu$ and problem parameters $\bpsi$. We discuss our results in the following section.

\section{Results}
\label{S:results}

\begin{figure*}
\includegraphics[width=\textwidth]{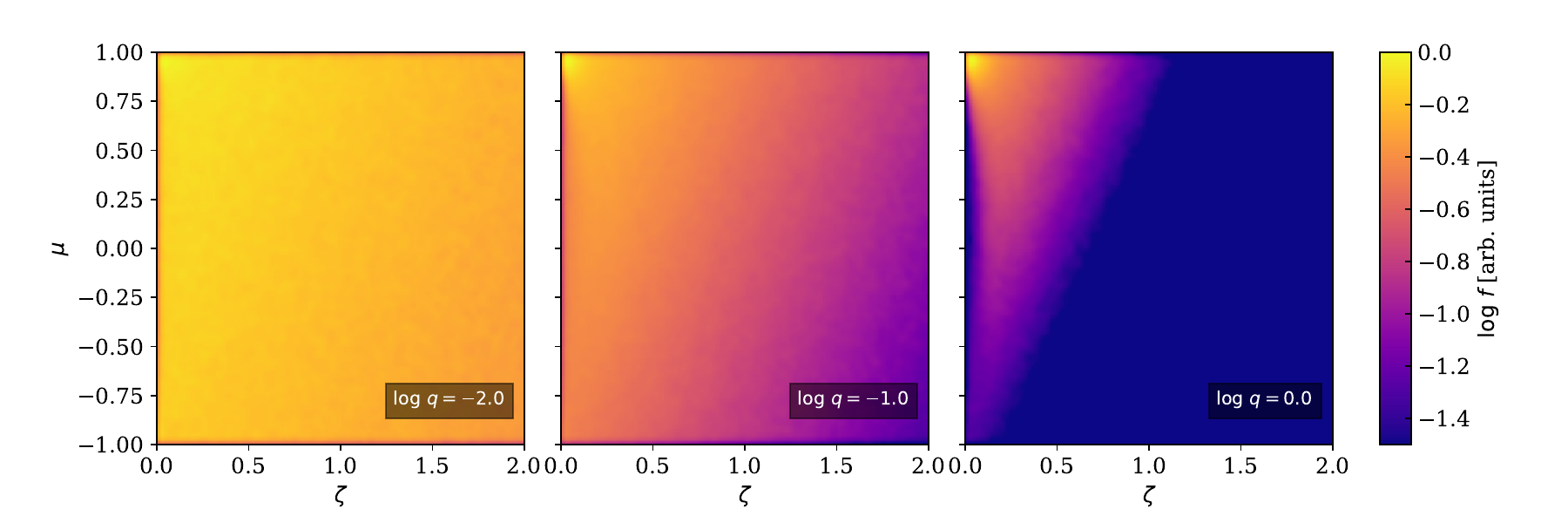}
\caption{Steady-state solutions to the CRe transport equation, showing the distribution function $f(\zeta, q, \mu)$ for our fiducial case ($k_p = -1.51$, $\log q_\mathrm{cut} = 1.6$, and $\mu_{\rm inj}=0.86$) as a function of position $\zeta$ and pitch angle $\mu$ for three different momenta $q$, as indicated in the labels of each panel. To facilitate comparisons across different values of $q$, the distribution functions in each panel have been normalized to set the maximum to unity.}
\label{f:CR_dist}
\end{figure*}

We have seen that the CRe distribution as a function of position $\zeta$ depends on both the properties of the CRe injection source -- as characterised by the spectral index $k_p$, cutoff momentum $q_\mathrm{cut}$, and injection opening angle $\mu_\mathrm{inj}$ -- and on the pitch angle $\mu$ and energy $q$ at which we measure. Consequently, the observable emission from this population will depend on the source properties, the orientation of the magnetic field relative to the observer (which determines the CRe pitch angle $\mu$ that a given observer will be able to see), and the energy at which the observation is made (which roughly determines the CRe momentum $q$ that contributes most to the observable emission). Thus our metrics of interest (as defined in \autoref{SS:metrics}) are defined over a five-dimensional parameter space $(k_p, q_\mathrm{cut}, \mu_\mathrm{inj}, q, \mu)$. Our ultimate goal in this section is to map out where in this five-dimensional space observable displaced emission is likely to arise. To reach this goal, in \autoref{SS:case_study} we first walk through the steps of our analysis for one particular combination of source properties $(k_p, q_\mathrm{cut}, \mu_\mathrm{inj})$, in order to build intuition for how the observable emission varies with $q$ and $\mu$ for a fixed source. Armed with this understanding, we then carry out a full study of the parameter space in \autoref{SS:parameter_space}.

\subsection{Case Study}
\label{SS:case_study}

\begin{figure}
\includegraphics[width=\columnwidth]{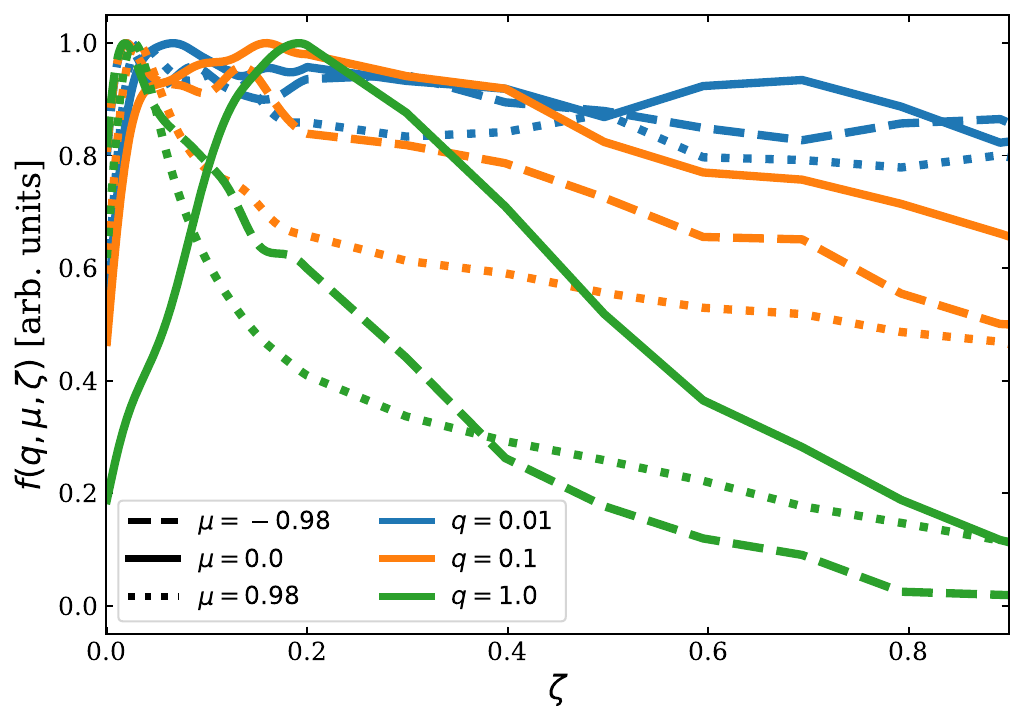}
\caption{CRe distributions for our fiducial case ($k_p = -1.51$, $\log q_\mathrm{cut} = 1.6$, and $\mu_{\rm inj}=0.86$) $f(\mu, q, \zeta)$, normalized by its maximum, shown for three pitch angles $\mu = -0.98, 0, 0.98$ (different line styles) at two momenta. Low-$q$ distributions are broad and nearly isotropic, whereas high-$q$ distributions are narrower and exhibit a clear $\mu$ dependence, with the largest peak displacement for $\mu = 0.0$. The values of our metrics of interest are given in the following \autoref{t:table}, with further discussion provided in \autoref{SS:case_study}.}
\label{f:1Ddist}
\end{figure}

%Normalized cosmic ray (CR) spatial distributions as a function of position for an injection angle of $\mu_{\rm inj} = 0.9$, shown for three different momenta (colored curves). The peak position of the distribution remains nearly constant across momenta, while the skewness increases with the power-law index $q$. \\ \textit{Right:} Ratio of the anisotropic CR distribution to the isotropic case as a function of pitch angle $\mu$, highlighting the enhanced brightness for lower $q$ values. Higher $q$ produces a more skewed but dimmer distribution. Both the injection angle and particle spectral index $k_p$ have only minor effects on the overall shape, though a shallower $k_p$ leads to higher brightness.}
%\label{f:dist}
%\end{figure}

\begin{figure}
\includegraphics[width=0.9\columnwidth]{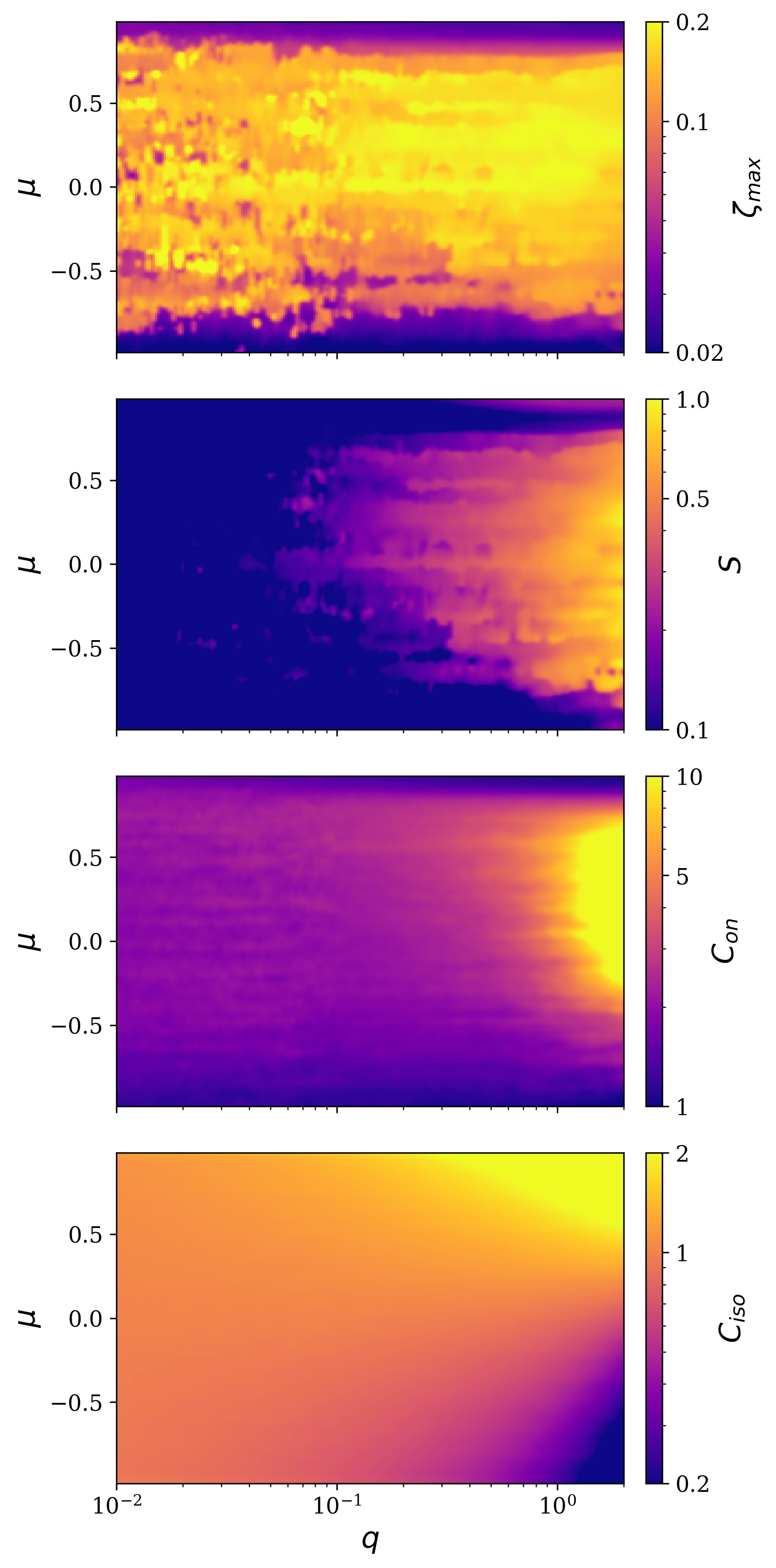}
\caption{
Four metrics of interest for detectable displaced CRe signals for our fiducial case ($k_p = -1.51$, $\log q_\mathrm{cut} = 1.6$, and $\mu_{\rm inj}=0.86$) as a function of dimensionless CRe momentum $q$ and the cosine of the angle between the magnetic field and the observer $\mu$. 
\textit{Top Panel:} The dimensionless position of the maximum of the signal, $\zeta_\mathrm{max}$.
\textit{Second Panel:} Sharpness parameter $S$ measuring the ratio of the distance from source to the peak of the emission $\zeta_\mathrm{max}$ to the FWHM of the emission.
\textit{Third Panel:} $C_\mathrm{on}$, defined as the ratio of the signal brightness at its (displaced) maximum to the signal brightness at the position of the CRe source.
\textit{Bottom Panel:} Ratio of the brightness at the emission peak to the brightness that would be produced by a source emitting cosmic rays with an isotropic distribution of pitch angles, $C_\mathrm{iso}$. Note that the dynamic range on the colour bar is the same for all panels. Each panel is centered on its characteristic mid-value -- 
0.1 for $\zeta_\mathrm{max}$, 
0.5 for $S$, 
5 for $C_{\rm on}$, 
and 1 for $C_{\rm iso}$ -- 
with logarithmic scaling spanning one decade in dynamic range 
($0.1\times v_{\mathrm{max}}$ to $v_{\mathrm{max}}$), 
where $v_{\mathrm{max}} = [0.2,\, 1.0,\, 10,\, 2.0]$. 
}
\label{f:color_map_2}
\end{figure}

\begin{table}
%\centering
%\begin{center}
\begin{tabular}{||c| c| c| c| c||} 
\hline
$q$ & $\mu$ & $\zeta_{\max}$ & $S$ & $C_{\text{on}}$ \\
\hline
     & $-0.98$ & 0.020 & 0.003 & 1.249 \\
0.01 &  $\phantom{-}0.00$ & 0.066 & 0.009 & 1.955 \\
     &  $\phantom{-}0.98$ & 0.034 & 0.006 & 1.791 \\
\hline
     & $-0.98$ & 0.020 & 0.011 & 1.240 \\
0.10 &  $\phantom{-}0.00$ & 0.158 & 0.066 & 2.132 \\
     &  $\phantom{-}0.98$ & 0.030 & 0.021 & 1.694 \\
\hline
     & $-0.98$ & 0.018 & 0.037 & 1.227 \\
1.00 &  $\phantom{-}0.00$ & 0.192 & 0.304 & 5.321 \\
     &  $\phantom{-}0.98$ & 0.026 & 0.113 & 1.618 \\
\hline
\end{tabular}
\caption{Values of the metrics of interest for the distributions plotted in \autoref{f:1Ddist}. Further discussion about these metrics are provided in \autoref{SS:case_study}.}
\label{t:table}
%\end{center}
\end{table}

To start our analysis, we examine a single sample value of the parameters describing the source, $(k_p, q_\mathrm{cut}, \mu_{\rm inj})$. For our example we choose the values that \citet{Krumholz2024} find provide the best fit for Terzan 5; these are $k_p = -1.51$, $\log q_\mathrm{cut} = 1.6$, and $\mu_{\rm inj}=0.86$.

\autoref{f:CR_dist} presents 2D slices through the steady-state distribution function $f(\zeta, q, \mu)$ for this set of source parameters at three different values of momentum $q$, allowing us to see how the distribution as a function of position $\zeta$ and pitch angle $\mu$ changes for different momenta. For high-momentum CRe ($q \gtrsim 1$) and forward-directed pitch angles ($\mu \gtrsim 0$), the distribution peak shifts toward positive $\zeta$, producing the displacement away from the source at $\zeta=0$ that is our main interest in this study. By contrast, at low momenta ($q \ll 1$), the distribution broadens and the peak is both much less sharp and shows minimal angular dependence. This behaviour is rooted in synchrotron cooling: the cooling time scales inversely with momentum ($t_{\rm cool} \propto 1/q$), so high-energy CRs cool before they can isotropize in pitch angle and thus retain the injection anisotropy, leading to displaced distributions. By contrast, low-energy CRe isotropize before cooling, yielding symmetric emission centred on the injection site. This mechanism naturally explains why high-energy non-thermal emission often appears spatially offset from the source. 

To build on our intuition more, we show the normalized (by the maximum value of the distribution) CRe distribution as a function of position $\zeta$ for selected values of $q$ and $\mu$ in \autoref{f:1Ddist}, again for our example set of parameters $(k_p, q_\mathrm{cut}, \mu_\mathrm{inj})$. The curves shown in this figure are effectively horizontal cuts through the three panels in \autoref{f:CR_dist} at the bottom, middle, and top of each box. We report the values of our metrics of interest for each of the curves shown in this figure in \autoref{t:table}. The figure shows that the peak location of the distribution is close to zero for $\mu = \pm 0.98$ and for $q = 0.01$, but that it is noticeably displaced at $\mu = 0$ and $q = 0.1$ and 1.0. The sharpness and shape of the distribution are also sensitive to $q$. Higher values of $q$ yield distributions that are more sharply peaked, and where the value of the distribution at $\zeta = 0$ is much smaller than the maximum value, whereas lower values produce flatter and broader profiles similar to an isotropic distribution. The numerical results reported in \autoref{t:table} demonstrate that our quantitative metrics capture the qualitative features visible in \autoref{f:1Ddist}. For the example cases shown, the value of $\zeta_\mathrm{max}$ is noticeably larger than zero only for $\mu = 0$ and $q = 0.1$ and $1.0$. Similarly, the sharpness parameter $S$ and ratio of peak to on-source values $C_\mathrm{on}$ also increase with increasing $q$ and toward $\mu = 0$ and away from $\mu = \pm -0.98$.

To further explore how our metrics vary with $q$ and $\mu$ for our example case, we plot them in \autoref{f:color_map_2}. The top panel shows $\zeta_\mathrm{max}$, and demonstrates that there is relatively little variation in its value across the parameter space: the peak is always at $\zeta_\mathrm{max} \approx 0.15 - 0.2$ for $q\gtrsim 0.1$ and $|\mu| \lesssim 0.7$. By contrast, the other parameters show far more variation: the sharpness parameter $S$ (second panel) and contrast with on-source emission $C_\mathrm{on}$ (third panel) are more than an order of magnitude larger at $q\sim 1$ and $\mu \sim 0-0.8$ than at $q\sim 0.1$ and any value of $\mu$ or at $q\sim 1$ and $\mu \lesssim -0.5$. This confirms the intuition we developed from \autoref{f:CR_dist} that displaced signals are likely to be most visible at high momentum and forward-biased pitch angles, while at lower momentum the signal blurs and the contrast between on- and off-source emission drops.

Finally, we see that $C_\mathrm{iso}$ (bottom panel) is relatively close to unity at all $q \lesssim 0.1$, which is not surprising: at low $q$, the cooling time is much longer than the isotropisation time, and thus CRs approach the results expected for an isotropic distribution. For $q\gtrsim 1$ the value of $C_\mathrm{iso}$ becomes strongly pitch angle-dependent, with enhanced emission close to $\mu = 1$ where the viewing angle is within the initial pitch angle distribution of which CRs are injected. By contrast, visible emission is strongly suppressed at $\mu < 0$, where CRs cool before having time to be scattered into the observers line of sight.

Examining the full set of metrics shown in \autoref{f:color_map_2}, we can draw the broad qualitative conclusion that, for our example set of source parameters, visibly-displaced emission is likely to be found if the emission traces CRs with $q\gtrsim 0.1$ and if the viewing angle is in the range $\mu \sim 0 - 0.8$. Smaller values of $q$ do not produce sharp emission peaks with significant contrast relative to the position of the source, while at smaller values of $\mu$ the emission is displaced but may be too dim to detect.

\subsection{Study of the Parameter Space}
\label{SS:parameter_space}

\begin{figure}
\includegraphics[width=\linewidth]{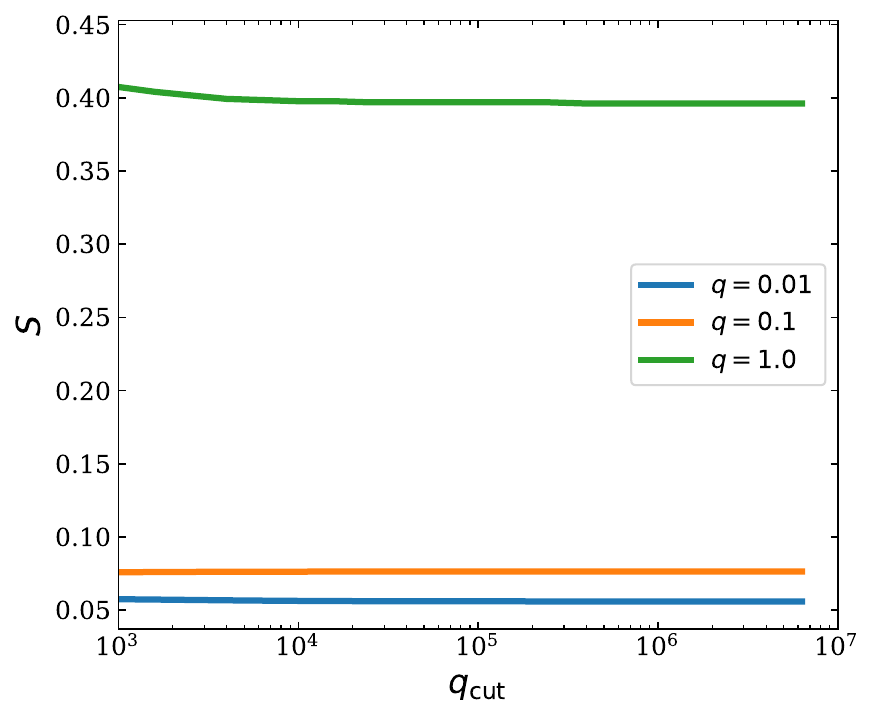}
\caption{Sharpness parameter $S$ as a function of $q_\mathrm{cut}$ for $\mu_\mathrm{inj} = 0.86$, $k_p = -1.51$ (as used in the case study in \autoref{SS:case_study}), and for an observer at $\mu=0.5$ and three CRe energies $q = 0.01, 0.1$, and 1. The figure illustrates that $q_\mathrm{cut}$ has no effect on $S$ as long as $q_\mathrm{cut} \gg q$.}
\label{f:par_S}
\end{figure}

Having understood the results from our sample model (spectral index of $k_p = -1.51$, injection pitch angle of $\mu_{\rm inj} = 0.86$, and cutoff parameter $\log q_{\rm cut} = 1.6$), we now extend our search of parameter space to examine how variations in $q_{\rm cut}$, $k_p$, and $\mu_{\rm inj}$ influence the resulting CRe distribution. In particular, we are interested in understanding what combination of source properties ($k_p$, $\mu_\mathrm{inj}$, $q_\mathrm{cut}$) and observing energy ($q$) and geometry ($\mu$) are likely to produce displaced emission. This is a five-dimensional parameter space, and thus challenging to survey thoroughly. However, we start by noting that we can eliminate one of the parameters: intuitively, we do not expect the cutoff energy $q_\mathrm{cut}$ to have much influence on the observable properties of the emission as long as it is much larger than the observing energy $q$. We verify this intuition for one particular example case in \autoref{f:par_S}, which shows the variation of $S$ with $q_\mathrm{cut}$ for an example set of parameters. As expected, we see almost no effect as long as $q_\mathrm{cut} \gg q$, and we will therefore only consider the limit $q_\mathrm{cut} \gg q$; this reduces our search from five dimensions to four.

For the remaining parameters, we compute all our metrics of interest on a grid of $k_p$ from $-3$ to $-1$ in steps of 0.2, of $\mu_\mathrm{inj}$ from $-0.5$ to 1 in steps of 0.1, of $\mu$ from $-0.9$ to $0.9$ in steps of 0.1, and in $\log q$ from $-2.0$ to $1.0$ (in steps of 0.5). We also supplement this coarse grid with a finer 2d grid of models at $\mu=0.5$, $q=1$, which has steps of 0.05 for $k_p$ and steps of 0.02 in $\mu_\mathrm{inj}$.

\subsubsection{Criteria for displaced emission}
\label{sssec:criteria}

\begin{figure}
\includegraphics[width=\linewidth]{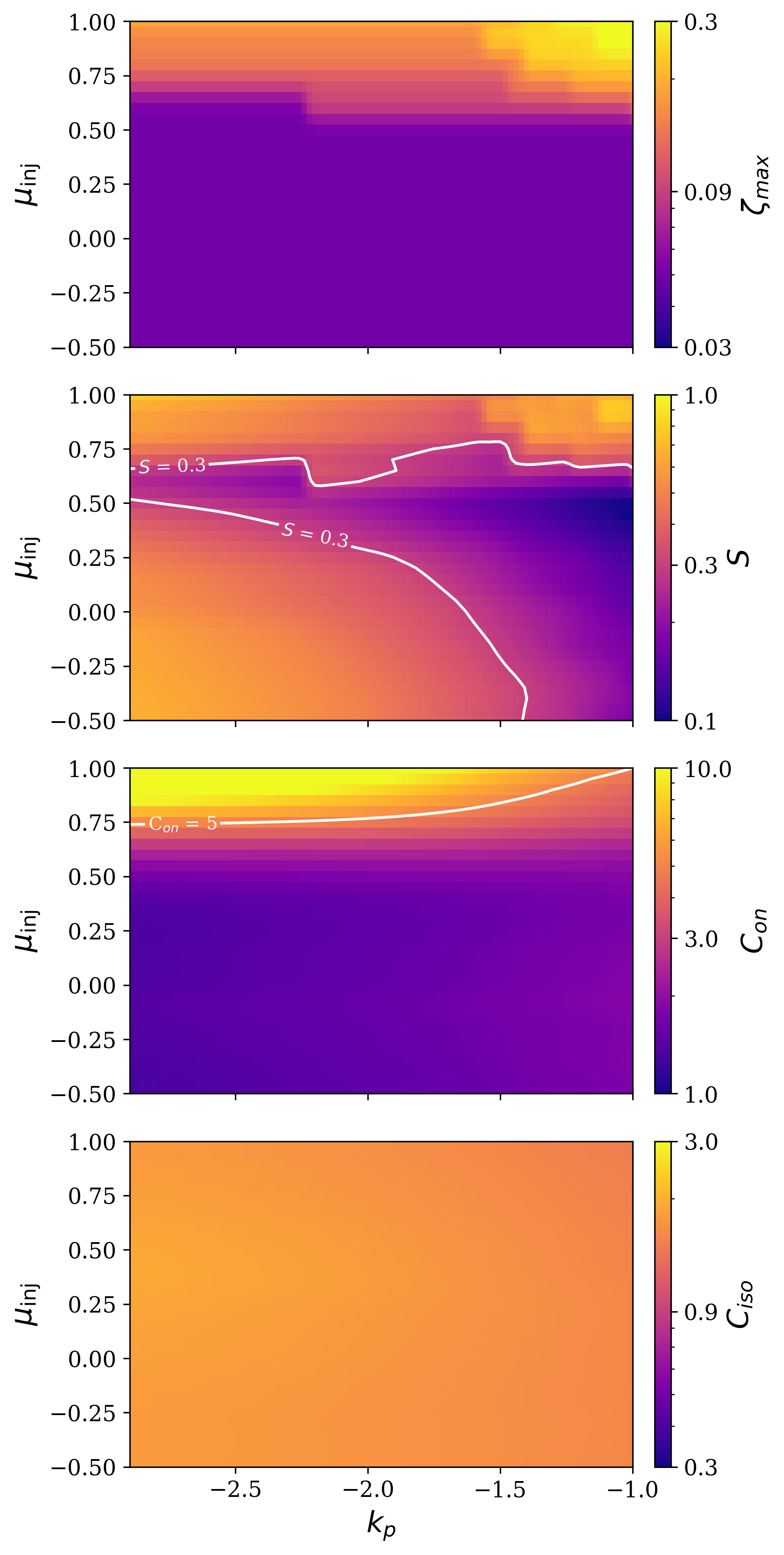}
\caption{Same as \autoref{f:color_map_2}, but now showing how the metrics of interest vary with $k_p$ and $\mu_{inj}$ for $\mu=0.5$ and $q=1$. The white contours in the middle two panels indicate $S = 0.3$ and $C_\mathrm{on} = 5$, our conditions for displaced emission.}
\label{f:par}
\end{figure}

%(in steps of 0.05 for the specific case of $\mu=0.5$ and $q=1$ and 0.2 for the corner plot), $\mu_{\rm inj}$ from $-0.5$ to $1.0$ (in steps of 0.02 for the specific case of $\mu=0.5$ and $q=1$ and 0.1 for the corner plot), $\mu$ from $-0.9$ to $0.9$ (in steps of 0.1), and $\log_{10}q$ from $-2.0$ to $1.0$ (in steps of 0.5).

%, and $q_{\rm cut}$ over the range $10^3$-$10^7$. 
We now investigate which points in the parameter space satisfy the conditions for showing displaced emission, as characterised by our metrics $S$, $C_\mathrm{on}$, and $C_\mathrm{iso}$. These are continuous variables, and which sources count as displaced will in the end always depend on the sensitivity of the measurement, but for the purpose of definiteness we say that a case shows displaced emission if it satisfies $S > 0.3$, $C_{\mathrm{on}} > 5$, and $C_{\mathrm{iso}} > 0.3$ -- in words, the ratio of the displacement of the emission peak from the source to the width of the emission peak must be $>30\%$, emission at the peak location must be at least five times brighter than that on the source, and the total observable luminosity from our anisotropic must be no lower than 30\% of the luminosity of an isotropic source at the same CRe energy. 

We illustrate the application of these cuts to our higher-resolution 2d grid of models at $q=1$, $\mu=0.5$ in \autoref{f:par}, which shows how each of four metrics of interest vary with $k_p$ and $\mu_\mathrm{inj}$ at this $(q,\mu)$ pair. The bottom panel shows that $C_{\mathrm{iso}}$ depends only weakly of $k_p$ and $\mu_\mathrm{inj}$, and is relatively close to unity everywhere; thus all points in this 2d plane satisfy the condition $C_\mathrm{iso} > 0.3$. On the other hand, this is not true of the other conditions. The second panel shows that the requirement that $S>0.3$ excludes a large part of parameter space at higher $k_p$ and smaller $\mu_\mathrm{inj}$, while the third panel shows that $C_{\mathrm{on}}>5$ is satisfied only for $\mu_{\mathrm{inj}} \gtrsim 0.75$ and $k_p \lesssim -1$. This is the most restrictive of our three conditions for this particular $(q,\mu)$ combination, and implies that for this momentum and viewing angle we only expect to see displaced emission from sources with narrow injection pitch angle distributions ($\mu_\mathrm{inj} \gtrsim 0.75$) and injection spectra that are not too shallow ($k_p \lesssim -1.5$). A final point worth noting is that, within the permitted region of parameter space, $\zeta_{\max}$ varies by less than a factor of two, as demonstrated in the top panel of \autoref{f:par}. Thus if we do see displaced emission for this particular $(q,\mu)$ combination, we can fairly immediately deduce the corresponding CRe diffusion pitch angle coefficient from the displacement distance, since the ratio of the physical displacement distance to the dimensionless $\zeta_\mathrm{max}$ depends on $K_\mu$ alone. We shall return to this point momentarily. 
%\mrtext{However, if $K_\mu$ itself depends on $q$, this inference would require modification; we defer a detailed exploration of such a dependence to future work.}
{In the more general case where $K_\mu$ depends on CR energy, the value we deduce will correspond to that at the energy that dominates the observed emission; however, we defer further exploration of this case to future work.}

\subsubsection{Parameters favourable to detecting displaced emission}
\label{SS:parameter}
\begin{figure*}
\includegraphics[width=\linewidth]{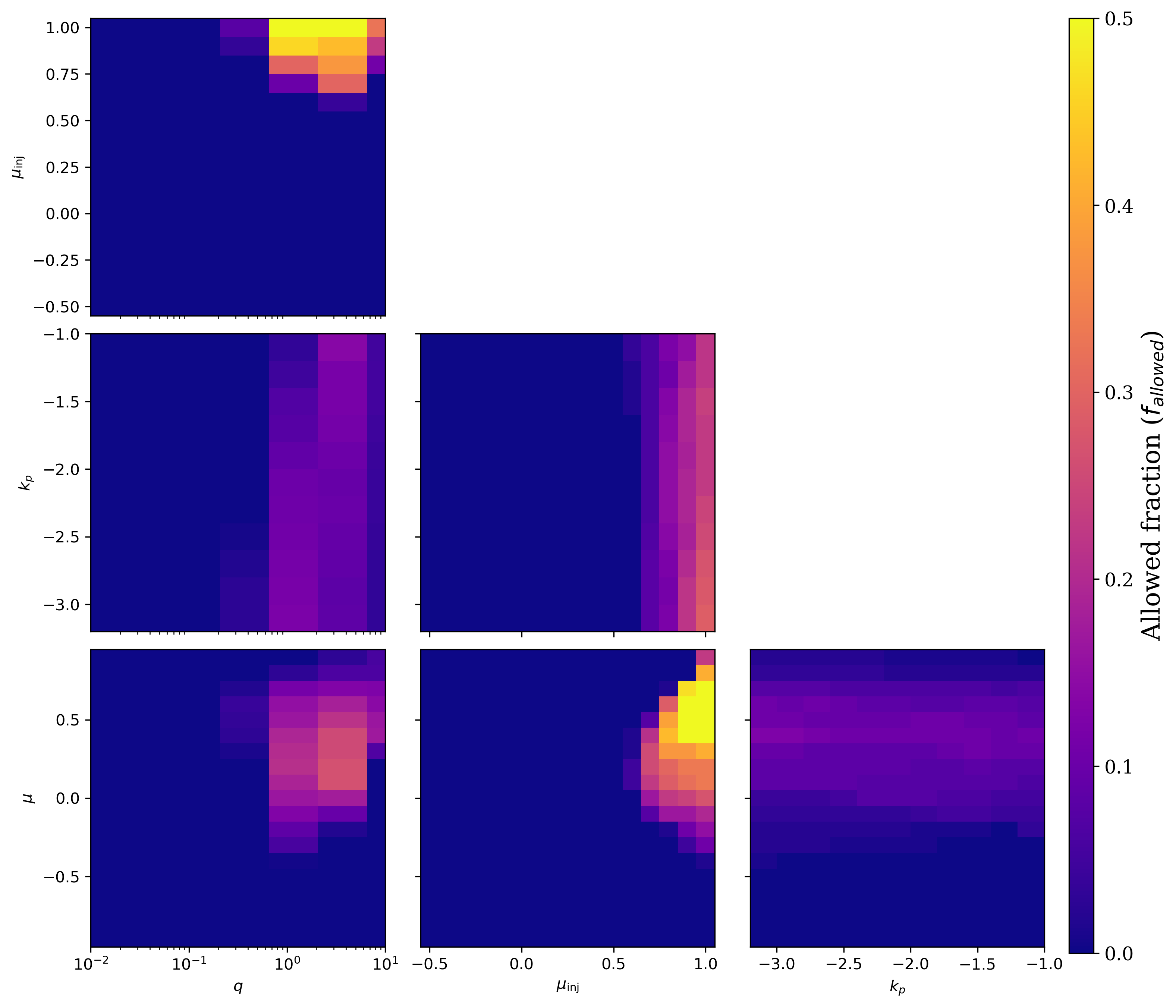}
\caption{Corner-style heatmaps showing, for each parameter pair, the fraction of models in our coarse grid (see \autoref{SS:parameter_space}) that are ``allowed'', meaning that they satisfy all of our criteria for measurable displaced emission. The colour indicates the allowed fraction from $0$ (no allowed models) to $1$ (all models allowed).\\}
\label{f:par_2}
\end{figure*}

To visualise where in our broader four-dimensional parameter space of $(k_p, \mu_\mathrm{inj}, \mu, q)$ displaced emission is likely to occur, in \autoref{f:par_2}, we show a corner-style heatmap showing, for each pair of model parameters, the fraction of models in our 4d coarse grid at that pair that meet our criteria for displaced emission. We refer to models that meet these criteria as ``allowed'', and for a fixed pair of parameters $(p_A, p_B)$ we define the allowed fraction as
\[
f_{\rm allowed}(p_A, p_B) = 
\frac{\text{\# of allowed models at } (p_A,p_B)}
     {\text{Total \# of models at } (p_A,p_B)}.
\]
The value of $f_{\rm allowed}$ ranges continuously from $0$ (no model allowed at that particular pair of parameters) to $1$ (all models allowed). Since the full model space is four-dimensional, each panel in \autoref{f:par_2} represents a projection onto a specific two-dimensional slice. In this projection, regions with high allowed fractions indicate that, for that pair of parameters, many combinations of the remaining parameters yield measurable displaced emission, while regions with low fractions indicate that few or no combinations of the other parameter do.

The figure demonstrates that displaced emission is restricted to CRs with energies corresponding to $\log q \gtrsim -0.5$ produced by sources with forward-peaked injection angles, $\mu_{\rm inj} \gtrsim 0.7$. The constraint on the viewing angle is more relaxed, with acceptable solutions spanning $\mu \in [-0.4,\, 0.8]$. The admissible spectral-index range is even broader with allowed models spanning the full range of our grid, $k_p \in [-3.1,\,-1.0]$

%We illustrate these trends using three representative values:  $\log_{10}q = 0.5$, $\log_{10}q = 0.0$, and $\log_{10}q = 1.0$. For $\log_{10}q = 0.5$, only a narrow portion of the parameter space satisfies the criteria. Acceptable models require strongly forward-peaked injection angles, $\mu_{\rm inj} \in [0.9,\, 1.0]$, and steep spectral indices, $k_p \in [-3.1,\,-2.5]$. The line-of-sight angle is similarly restricted to a relatively small interval, $\mu \in [0.3,\, 0.7]$. For $\log_{10}q = 0.0$, the allowed region broadens significantly. Injection angles down to $\mu_{\rm inj} \approx 0.7$ become permissible, and the admissible spectral-index range expands to $k_p \in [-3.1,\,-0.9]$. The constraint on the viewing angle is also greatly relaxed, with acceptable solutions spanning $\mu \in [-0.4,\, 0.8]$. For $\log_{10}q = 1.0$, all forward-peaked injection angles with $\mu_{\rm inj} \gtrsim 0.7$ satisfy the criteria, and even relatively flat spectra are allowed, extending to $k_p \in [-3.1,\,-0.1]$. However, the viewing-angle range shifts toward more forward orientations, $\mu \in [0.2,\, 0.9]$, indicating a more restrictive geometric requirement than in the $\log_{10}q = 0.0$ case.

\subsubsection{Displacement distances}

%Additionally, in \autoref{f:par_zeta_max}, we show the probability distribution function of   $\zeta_{\max}$ across the permitted region of parameter space. We find that $\zeta_{\max}$ only varies by less than a factor of $3$ across the allowed parameter space. It implies that the displacement of the signal, $\zeta_{\max}$, is largely insensitive to our parameters of interest. 

\begin{figure}
\includegraphics[width=\linewidth]{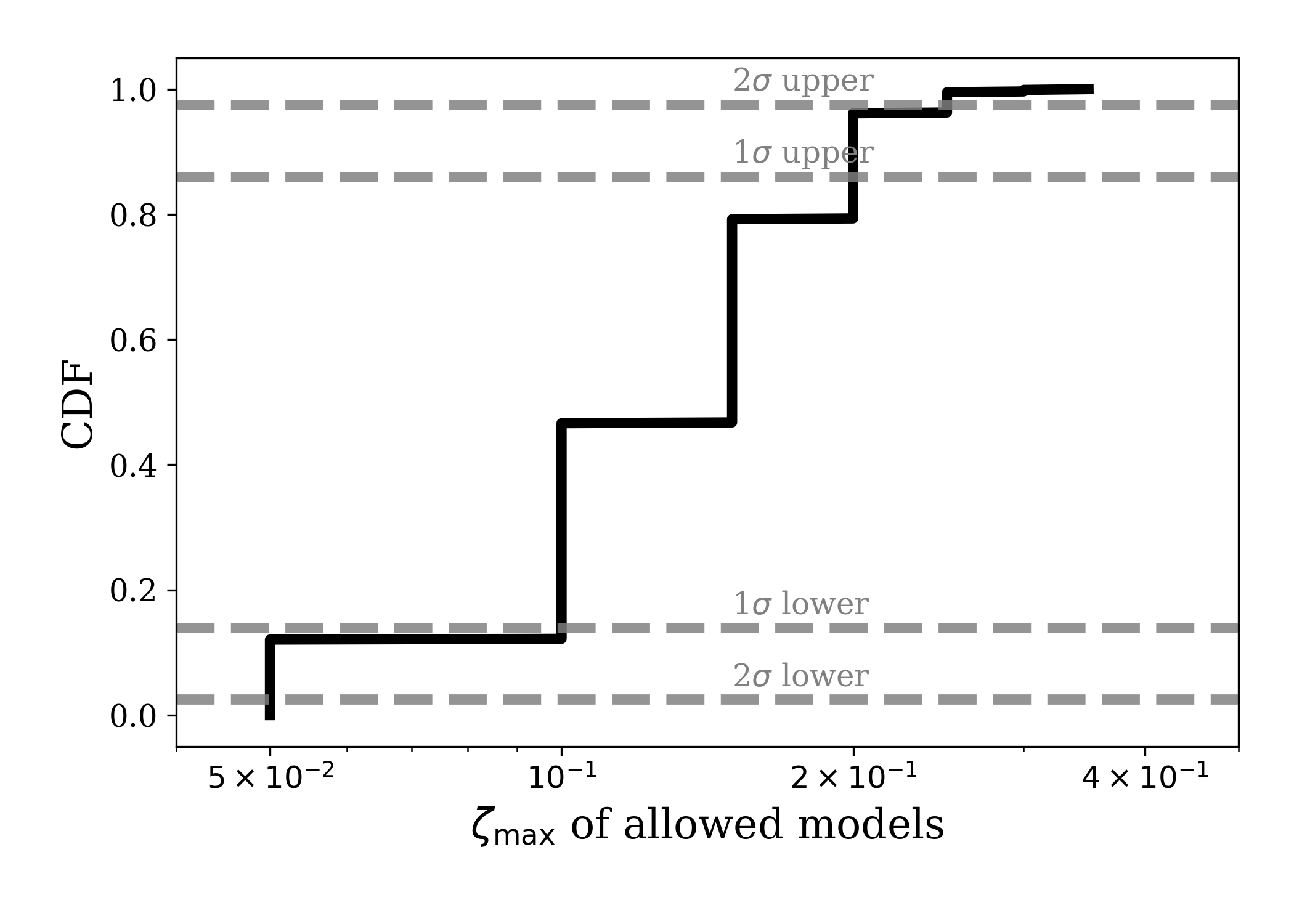}
\caption{Cumulative distribution function (CDF) of $\zeta_{\rm max}$ for all ``allowed'' models (i.e., models where we expect observable displaced non-thermal emission) in our grid.}
\label{f:par_zeta_max}
\end{figure}

We note that, for the example explored in \autoref{sssec:criteria}, the parameter that leads to detectable displacement of emission also leads to a fairly narrow range of displacement distance, $\zeta_\mathrm{max}$. In \autoref{f:par_zeta_max} we test whether this is a general phenomenon or restricted to that example case by plotting the cumulative distribution function (CDF) of $\zeta_{\max}$ across all allowed models in our grid. The dashed gray lines mark the 1$\sigma$ (16th to 84th percentile) and 2$\sigma$ (2.5th to 97.5th percentile) range. From the CDF, we find that $\zeta_{\max}$ lies within $0.10$–$0.20$ at 1$\sigma$, and $0.05$–$0.25$ at 2$\sigma$. This demonstrates that the trend we noted in \autoref{sssec:criteria} is more general: for ``allowed'' models that produce detectable displaced emission (by our criteria), $\zeta_\mathrm{max}$ values vary only by a factor of $\sim 2$ at the $2\sigma$ level, indicating a narrow and tightly constrained distribution. 
%This demonstrates that the displacement of the signal, $\zeta_{\max}$, is largely insensitive to variations in our model parameters, consistent with our earlier findings.

The narrow range of variation in $\zeta_\mathrm{max}$ is extremely convenient, because it means that, if we detect displaced emission from a source, we can obtain a rough estimate of $\zeta_\mathrm{max}$ without needing to carry out computationally expensive bespoke modelling to determine the full set of model parameters, as \citet{Krumholz2024} did for Terzan 5. Since the ratio of $\zeta_\mathrm{max}$ to the physical displacement distance depends only on $K_\mu$ and $c$, this in turn means that, for a source where we have measured the angular displacement of the emission and know the source distance, we can obtain a rough estimate of $K_\mu$ almost immediately. We will exploit this capability in \autoref{S:discussion}.

\section{Discussion}
\label{S:discussion}

Our results establish a clear framework for identifying the conditions under which displaced cosmic-ray (CR) emission arises and becomes observationally detectable. In particular, we delineate a well-defined region of parameter space, characterized by dimensionless particle momentum $q$, particle pitch angle $\mu$, and narrowness of the pitch angle distribution at the source $\mu_\mathrm{inj}$, where all three criteria for detection—localization, significant displacement, and sufficient brightness—are simultaneously satisfied. This region corresponds to an observational window of $\log q\gtrsim -0.5$, $-0.4 \lesssim \mu \lesssim 0.8$, and $\mu_\mathrm{inj} \gtrsim 0.7$. In words, we expect to observe displaced emission in situations 
where the photons we observe are being produced by CRs whose radiative loss time is at most a factor of $\sim 2-3$ larger than their pitch angle scattering time, when the accelerator producing the CRs injects them with a range of pitch angles $\lesssim 45^\circ$, and when the observer's viewing angle relative to the magnetic field is in a relatively broad range from $\approx 40^\circ - 110^\circ$. When displaced emission is detectable, the dimensionless distance $\zeta_\mathrm{max}$ between the acceleration site and the maximum of the observable emission falls into a factor of $\sim 2$ range around $0.1$-$0.15$. Here we discuss some important implications of these results.

%We find that variations in the cutoff parameter $q_{\rm cut}$ have little impact on the extent of this allowed parameter space. By contrast, changes in the spectral index $k_p$ and the injection pitch angle $\mu_{\rm inj}$ can modify the diagnostic metrics by up to a factor of four across the explored ranges. A narrow injection angle enhances the metric values, thereby favoring detectability. In addition, steeper spectra lead to a marked increase in $C_{\rm on}$ and $C_{\rm iso}$, further strengthening the displaced signal.

\subsection{Pitch angle scattering rates for known and future displaced sources}
\label{ss:sc_rate}
A particularly noteworthy implication of our results is that, in situations where displaced emission is likely to be detectable, the location of the emission peak is largely insensitive to either the CRe momentum (and thus the observing energy) or to the properties of the source. This means that, when displacement is observed, it provides a direct diagnostic of the pitch-angle scattering rate. A similar observation was recently made by \citet{Cao2025}, whose Figure 2 shows that the offset between the $\gamma$-ray source 1LHAASO J1740+0948u and its associated pulsar PSR J1740+1000 remains nearly constant across the observed photon energy range ($\log (E_\gamma/\rm TeV) = 1.4$–$2.4$). In this context, the displacement distance $z_\mathrm{max}$ serves as a proxy for the pitch-angle scattering rate:
\begin{equation}
    K_\mu = c \left(\frac{\zeta_\mathrm{max}}{z_\mathrm{max}}\right) \sim 10^{-9} \left(\frac{z_\mathrm{max}}{\mathrm{pc}}\right)^{-1}\,\left(\frac{\zeta_\mathrm{max}}{\mathrm{0.1}}\right)\,\mathrm{s}^{-1},
    \label{eq:Kmu_approx}
\end{equation}
where in the numerical evaluation we have used $\zeta_\mathrm{max} \approx 0.1$, as found by our numerical experiments. The corresponding spatial diffusion coefficient is
\begin{equation}
    K_x = \frac{c^2}{6 K_\mu} \sim 10^{29} \left(\frac{z_\mathrm{max}}{\mathrm{pc}}\right) \, \mathrm{cm}^{2}\,\mathrm{s}^{-1}
    \label{eq:Kx_approx}
\end{equation}
We caution that this is only a rough estimate -- at a minimum, it contains a geometric uncertainty because the observable quantity is not $z_\mathrm{max}$ but $|z_\mathrm{max} \mu|$, the projected displacement on the sky -- and a more precise estimate can be achieved by a full fit to the joint spatial and spectral distribution, as \citet{Krumholz2024} carried out for Terzan 5. Nonetheless, \autoref{eq:Kmu_approx} and \autoref{eq:Kx_approx} can be used to give a quick first estimate of the pitch angle and spatial diffusion coefficients with no need for additional modelling. 
%For example, a displacement of $x$ parsecs corresponds to a pitch-angle diffusion coefficient of order $K_\mu \sim c/x \approx 10^{-9}\,\mathrm{s}^{-1} \left(\frac{x}{10\,\mathrm{pc}}\right)^{-1}$.

This insight has broader relevance for interpreting displaced emission features in various astrophysical systems -- such as $\gamma$-ray halos around pulsars or radio relics in galaxy clusters -- where CRs are likely injected anisotropically and subject to energy-dependent scattering. In such cases, the displacement of the emission peak may offer a powerful observational constraint on the microphysics of CR transport, such as scattering rate.

%In summary, the emergence of observable displaced CR emission hinges on a delicate interplay between transport physics and injection geometry. Only under conditions of narrow pitch-angle injection, moderate particle momentum, and favorable viewing angle does the CR distribution produce a displaced signal that is both detectable and spatially distinct.

\subsection{Why are there no displaced GeV $\gamma$-ray sources?}
\label{ssec:why_no_GeV}

A further implication of our analysis is that we can explain an otherwise-puzzling aspect of the observational phenomenology of displaced non-thermal emission: while displacement has been seen in both very high to ultra high energy (TeV-PeV) $\gamma$ rays (by HESS, HAWC, and LHASSO) and X-rays (by \textit{Chandra} and \textit{XMM-Newton}), it has \textit{not} been observed in GeV $\gamma$-rays by \textit{Fermi}. Given \textit{Fermi}'s all-sky coverage and high sensitivity compared to the TeV detectors, the absence of displaced \textit{Fermi} sources suggests that this is not merely an observational bias, but a real physical effect: displaced emission does not appear to occur at GeV energy at the same rate as at TeV energies, or in X-rays.

We can understand this effect as arising from the requirements we have obtained for displaced emission. First, we have seen that displaced emission requires $q\gtrsim 1$; returning from dimensionless to dimensional variables (\autoref{SS:numerics}), this is equivalent to requiring that the CRe momentum obey
\begin{equation}
    p \gtrsim K_\mu t_{c,0} m_e c \approx \frac{m_e^2 c^2 K_\mu}{\sigma_T U_B},
\end{equation}
where $K_\mu$ is the pitch angle scattering rate and $U_B$ is the magnetic energy density, and we have dropped factors of order unity since this is an order of magnitude argument. We can write this in a more useful form by eliminating $K_\mu$ in favour of the ratio of dimensionless to physical lengths, since we have seen that the dimensionless distance to the emission maximum $\zeta_\mathrm{max}$ is nearly fixed. Doing so, and multiplying $p$ by $c$ to constrain the electron energy $E_e$, gives
\begin{align}
    E_e & \gtrsim \frac{m_e^2 c^4}{\sigma_T U_B} \left(\frac{\zeta_\mathrm{max}}{z_\mathrm{max}}\right)  \nonumber \\
    & \approx 130\left(\frac{U_B}{10\,\mathrm{eV}\,\mathrm{cm}^{-3}}\right)^{-1} \left(\frac{z_\mathrm{max}}{10\,\mathrm{pc}}\right)^{-1}\,\mathrm{TeV},
    \label{eq:Ez}
\end{align}
where $z_\mathrm{max}$ is the distance between the acceleration site and the emission maximum in physical length units, and for the purposes of numerical evaluation we have substituted $\zeta_\mathrm{max} \approx 0.1$.

The implication of this expression is that, if the magnetic field causing the energy loss has an energy density of order 10 eV cm$^{-3}$ (somewhat higher than the mean of the ISM in the Solar neighbourhood, but comparable to what might be expected in shock-compressed regions), and the displacement between the accelerator and the emission peak is of order 10 pc, then displaced emission will only be produced by CRs with energy $E_e$ greater than tens of TeV. Conversely, for displaced emission to be produced by CRe with energies of tens of GeV rather than tens of TeV, the magnetic energy density would need to approach 10 keV cm$^{-3}$ (corresponding to field strengths in the mG range) or the displacement distance would have to increase to kpc scales. The latter possibility we can rule out both because it is implausible that the magnetic fields confining the CR would remain relatively straight over such long distances, and because if the field did remain straight, the angular distance between the accelerator and the emission would be so large that it would not be possible to associate the two. The former option -- mG strength fields -- is not impossible, but requires field strengths much, much larger than typical interstellar magnetic field strengths on pc or larger scales.

These constraints on the energies of the CRs generating the emission in turn limit the photon energies at which displaced emission is likely to be seen. For an electron of energy $E_e \sim 10-100$ TeV, we can consider three main emission processes: inverse Compton scattering of starlight photons ($E_\gamma \sim \mathrm{eV}$), inverse Compton scattering of photons from the cosmic microwave background (CMB) or dust thermal emission ($E_\gamma \sim 10^{-4} - 10^{-3}$ eV), and synchrotron emission. For the two inverse Compton channels, we have $\Gamma_e = 4 E_\gamma E_e / m_e^2 c^4 \gg 1$ for starlight photons and $\lesssim 1$ for CMB or dust, placing the former process in the Klein-Nishina limit and the latter in the Thomson scattering regime \citep{Blumenthal70a}. In the Klein-Nishina limit the scattered photon energies are typically close to the initial photon energy, $\sim 10-100$ TeV, while in the Thomson limit the mean emitted photon energy is $(4/3) E_\gamma (E_e / m_e c^2)^2 \sim 1$ TeV. Finally, most synchrotron emission occurs around the critical energy
\begin{align}
    E_\mathrm{c,sy}  &= \frac{3 h e B}{4 \pi m_e c} \left(\frac{E_e}{m_e c^2}\right)^2  \nonumber\\ & \approx 13 \left(\frac{E_e}{100\,\mathrm{TeV}}\right)^2 \left(\frac{U_B}{10\,\mathrm{eV}\,\mathrm{cm}^{-3}}\right)^{1/2}\,\mathrm{keV},
    \label{eq:synch_crit}
\end{align}
placing this emission in the X-ray. We can therefore understand why displaced emission is a phenomenon seen in TeV $\gamma$-rays and X-rays but not in GeV $\gamma$-rays: for reasonable interstellar magnetic field strengths and for the range of displacements that is plausibly detectable, displaced emission will always come from leptons with energies in the tens to hundreds of TeV range, which in turn will produce TeV $\gamma$-ray emission via inverse Compton scattering and keV X-ray emission via synchrotron, but will not emit strongly at GeV energies.

\subsection{Can we observe displaced emission from other mechanisms or at other wavelengths?}
\label{ss:otherwavelength}
\autoref{ssec:why_no_GeV} provides an explanation for why systems like Terzan 5, where the main emission mechanisms are IC and synchrotron, show displacement at TeV and keV but not at GeV energies. However, we can further generalise these arguments to other radiation mechanisms and wavelengths, and ask under what circumstances displaced emission is possible more broadly -- for example, might we see displaced hadronic emission from a source that accelerates protons with an anisotropic pitch angle distribution, or displaced synchrotron emission in the radio rather than the X-ray?

To answer these questions, we begin by first recalling that a generic requirement for displaced emission is that particles should have a cooling time close to their scattering time.
In this case, the physical distance to their emission maximum approximately satisfies 
\begin{equation}
    z_{\rm max} = \frac{\zeta_{\rm max} c}{K_\mu} \simeq \zeta_{\rm max} c t_{\rm cool}
\label{eq:zmax}
\end{equation}
Using the first equality in \autoref{eq:Kx_approx},  
we can also infer the spatial diffusion coefficient that would be implied by a pitch angle scattering rate that generates an offset source given some $t_{\rm cool}$:
\begin{equation}
    K_x \sim \frac{c^2 t_{\rm cool}}{6} .
\label{eq:Kxroughly}
\end{equation}
Thus we can express both the characteristic displacement and the associated spatial diffusion coefficient in terms of the cooling time.

Second, consider that, for any given emission mechanism, there is a mapping between the energy of the emitting CRs $E_\mathrm{CR}$, the characteristic energies of the emitted photon $E_\gamma$, and the ambient ISM conditions. We can abstract this relationship as 
\begin{equation}
    E_{\rm CR} = E_{{\rm CR},m}(E_\gamma,\Theta_{\rm ISM}) 
\end{equation}
where $m$ refers to the specific emission mechanism (e.g., synchrotron, IC, bremsstrahlung, hadronic $pp$ emission) and $\Theta_{\rm ISM}$ is a vector of relevant ISM parameters,
\begin{equation}
\Theta_{\rm ISM} = \{E_{\gamma,i},u_{\rm rad}, B,n_\mathrm{H}\},
\end{equation}
where the important parameters for the mechanisms we have named might include one or more of the characteristic energy density $u_{\rm rad}$ and photon energy $E_{\gamma,i}$ of the interstellar radiation field (ISRF) component being up-scattered, the magnetic field amplitude $B$, and the ISM proton density $n_\mathrm{H}$.

Since the cooling time is also a function of the CR energy $E_\mathrm{CR}$, we can insert this dependence into \autoref{eq:zmax} and \autoref{eq:Kxroughly} to give 
\begin{equation}
    z_{\rm max}(E_\gamma,\Theta) \sim \zeta_{\rm max} c t_{\rm cool}(E_{{\rm CR},m}(E_\gamma,\Theta),\Theta) 
\label{eq:zmax2}
\end{equation}
and
\begin{equation}
    K_x(E_\gamma,\Theta) \sim \frac{c^2  t_{\rm cool}(E_{{\rm CR},m}(E_\gamma,\Theta),\Theta)}{6}.
\label{eq:Kinfrd}
\end{equation}
The meaning of these expressions is that we can express the displacement distance and spatial diffusion coefficient as a function of the emission mechanism $m$, the observed photon energy $E_\gamma$, and the vector of ambient ISM conditions $\Theta_\mathrm{ISM}$. These relationships allow us to diagnose where we are likely to be able to observe displacement for any emission mechanism.

The final ingredients we need to evaluate \autoref{eq:zmax2} and \autoref{eq:Kinfrd} are a value of $\zeta_\mathrm{max}$ -- we use $\zeta_\mathrm{max} = 0.1$ throughout -- along with expressions to map photon energy to CR energy for different emission mechanisms, and the cooling rate as a function of CR energy. For the photon to CR mapping, we adopt the approximate relationships \begin{eqnarray}
    E_{\rm CR,IC} &=&  
    \nonumber \\  1.5 E_\gamma + &0.44& \ {\rm TeV} \ 
    \sqrt{\frac{E_\gamma}{\rm TeV}\left(\frac{E_{\gamma,i}}{\rm eV}\right)^{-1} + \left(\frac{E_\gamma}{\rm TeV}\right)^2} 
    \\
    E_{\rm CR,sy} &=& 4.75 \ {\rm TeV} 
    \left(\frac{E_\gamma}{\rm eV}\right)^{1/2}
    \left(\frac{B}{\rm \mu G}\right)^{-1/2} \\
    E_{\rm CR,had} &=& 10 E_\gamma
\end{eqnarray}
for inverse Compton, synchrotron, and hadronic emission. In the first of these expressions, the two parts are designed to approximately capture the transition from the Thomson to the Klein-Nishina regime \citep{Blumenthal70a}. The second expression, for synchrotron, assumes the photon energy is equal to the critical energy and follows from a re-arrangement of \autoref{eq:synch_crit}. Finally, for the third expression we use the rule of thumb that hadronic collsions produce photons with typical energies $\approx 10\%$ of the initial CR energy \citep[e.g.,][]{Kafexhiu14a}. Finally, to compute the cooling timescale, we take into account ionisation \citep{Protheroe08a}, bremsstrahlung \citep{Baring99a}, IC, and synchrotron losses\footnote{However we caution the reader that, because we do not track the evolution of the pitch angle distribution in making these plots, they will tend to underestimate the synchrotron cooling time. This happens because we do not account for the  $\propto (1-\mu^2)^{-1}$ dependence on pitch angle in the synchrotron cooling time, important for electrons injected at small pitch angle. However, since we are attempting only an order-of-magnitude calculation we will omit this detail.} for CR electrons and $pp$ losses \citep{Kelner06a,Kafexhiu14a} for protons. For the purposes of computing the IC cooling rate, we approximate the ISRF as consisting of two monoenergetic components: the CMB ($E_{\gamma,i} = 6.3 \times 10^{-4}$ eV, $u_\mathrm{rad} = 0.26$ eV cm$^{-3}$) and starlight ($E_{\gamma,i} = 1$ eV, $u_\mathrm{rad} = 1$ eV cm$^{-3}$)\footnote{{Our fiducial, total radiation energy density of $u_\mathrm{rad} = 1.26$ eV cm$^{-3}$ would be in energy density equipartition with a magnetic field of amplitude 7.1 $\mu$G.}}, and unless otherwise specified we assume the ambient number density of H nuclei is $n_\mathrm{H} = 0.1$ cm$^{-3}$.

\begin{figure}
\includegraphics[width=\linewidth]{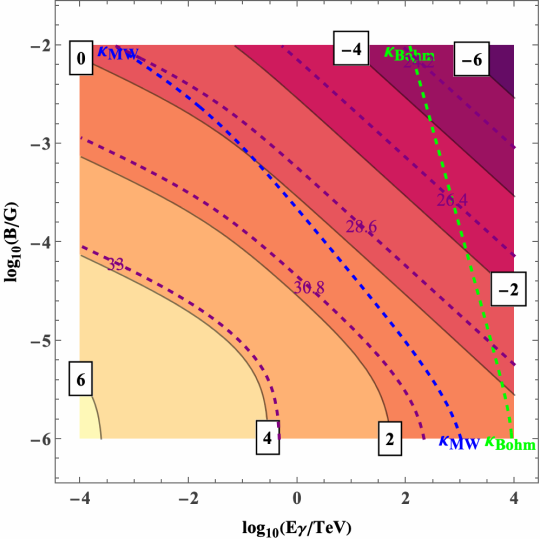}
\caption{Diagnostic plot for displaced emission produced by IC emission off optical/UV photons. The contours associated with the coloured background and labelled by the white background boxes show the log displacement distance $\log (z_{\rm max}/\mathrm{pc})$ (e.g., ``2'' corresponds to 100 pc; see \autoref{eq:zmax2}). The purple, dashed contours show the corresponding log spatial diffusion coefficient $\log (K_x/\mathrm{cm}^2\;\mathrm{s}^{-1})$ (\autoref{eq:Kinfrd}). The single blue contour show our empirical model for the typical CR diffusion coefficient in the Milky Way (\autoref{eq:kappaMW}), and the single green contour shows the Bohm limit diffusion coefficient (\autoref{eq:kappaBohm}).
}
\label{f:optIC}
\end{figure}

\begin{figure}
\includegraphics[width=\linewidth]{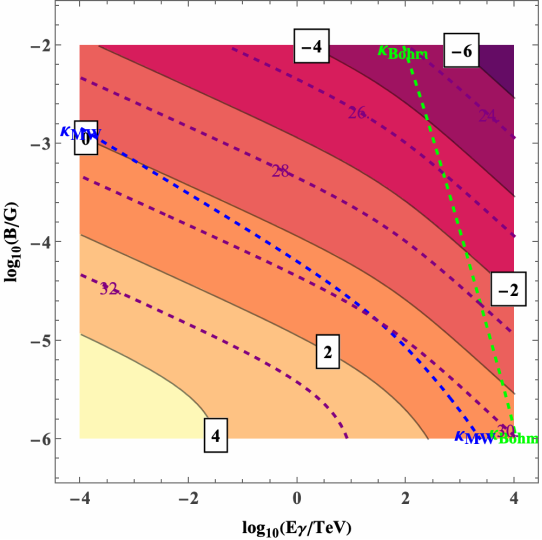}
\caption{Same as \autoref{f:optIC}, but for IC emission off CMB photons.}
\label{f:CMBIC}
\end{figure}

\begin{figure}
\includegraphics[width=\linewidth]{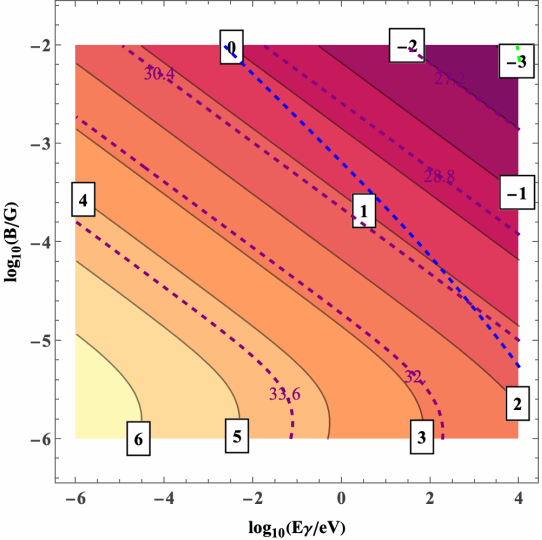}
\caption{Same as \autoref{f:optIC}, but for synchrotron  emission. Note that, unlike in \autoref{f:optIC}, here the $x$-axis for the emitted photon energy is in units of eV rather than TeV.}
\label{f:synch}
\end{figure}

\begin{figure}
\includegraphics[width=\linewidth]{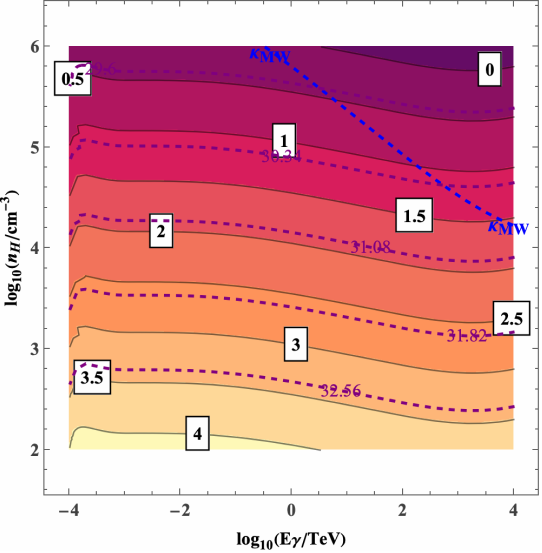}
\caption{Same as \autoref{f:optIC}, but for hadronic  emission off ISM gas. Note that the $y$-axis specifies the gas density, which is assumed to be constant at the given value around the accelerator.}
\label{f:hadron}
\end{figure}

In \autoref{f:optIC} to \autoref{f:hadron} we plot, in order, the maps that we obtain from this procedure for IC emission off optical / UV photons, IC emission off CMB photons, synchrotron emission, and hadronic emission, all as a function of observed photon energy $E_\gamma$ and ISM conditions -- $B$ for CR electrons, and $n_\mathrm{H}$ for CR protons. In these figures, solid contours labeled with white outlined numbers indicate $\log (z_\mathrm{max}/\mathrm{pc})$, while dashed purple contours indicate $K_x$. For comparison with the values of $K_x$, we also provide two other contours. The dashed blue contours shows a heuristic model for the energy-dependent spatial diffusion coefficient in the plane of the Milky Way,
\begin{equation}
    \kappa_{\rm MW}(E) = 3 \times 10^{28} {\rm cm^2 \ s^{-1}} \left(\frac{E_\mathrm{CR}}{\rm 4 \ GeV}\right)^{1/3}. 
\label{eq:kappaMW}
\end{equation}
This simple diffusion coefficient scaling, inspired by \citet{Evoli18a}, approximately reproduces the phenomenology presented by the local CR population (B/C ratio energy evolution, etc). The other comparison we provide is a dashed green contour corresponding to the Bohm diffusion coefficient
\begin{align}
    &\kappa_{\rm Bohm}(E_\mathrm{CR},B)=  \frac{c r_{\rm gyro}(E_\mathrm{CR},B)}{3} 
    \nonumber \\
    & \simeq  3.3 \times 10^{25} \ {\rm cm^2 \ s^{-1}} \left(\frac{E_\mathrm{CR}}{\rm TeV}\right)
    \left(\frac{B}{\rm \mu G}\right)^{-1},
\label{eq:kappaBohm}
\end{align}
which defines a lowest allowed spatial diffusion coefficient (corresponding to a maximum pitch angle scattering rate)\footnote{Note that both \autoref{eq:kappaMW} and
\autoref{eq:kappaBohm} are for the case of CR particles with charge $|Z|$ = 1 as we only consider either CR electrons or protons here, not, for simplicity, heavier ions.}.

Attending to \autoref{f:optIC} {\it et seq.} 
we can quickly come to some conclusions as to which environments and radiation mechanisms might be expected to allow pitch angle scattering to produce an observable offset source.
First, from \autoref{f:hadron}, we see that displaced hadronic emission due to the pitch angle anisotropy effect that is the focus of this paper is unlikely to occur in any nearby astrophysical environment: for pitch angle scattering rates anywhere close to those inferred for Galactic CRs, to produce a displacement of, say, $\sim$ 10 pc, one needs both an extremely high number density, $\sim 10^5$ cm$^{-3}$, and very high energy, $\gtrsim 100$ TeV, photons; gaseous structures of the lineal scale and density implied by this combination are either extremely rare or non-existent in the Milky Way, especially structures with a constant density as we have assumed here.
This leads to a more direct point: with hadronic emission, one may of course produce an ``offset'' source in the case of an inhomogeneous gas density distribution, but this is a  mechanism distinct from ours. 
The classic case here is a supernova remnant (SNR) accelerator offset from a dense molecular cloud which the escaping CRs illuminate \citep[e.g., the W28 SNR system;][]{HESS08a,Fermi10a}.
In passing, we find that the  situation for bremsstrahlung emission is similar to hadronic emission: the pitch angle scattering mechanism is not likely to lead to any offset source in any real Galactic environment\footnote{The diagnostic plot for bremsstrahlung is qualitatively similar to \autoref{f:hadron} but with the contours shifted down by half a dex; for brevity, we do not include it.}.

Working backwards, now consider the case of synchrotron emission shown in \autoref{f:synch}.
In general terms, for most Galactic environments,  synchrotron emission is overwhelmed by thermal emission over most of the electromagnetic spectrum over which it is emitted, except in radio continuum (RC; $\sim 10^{-6} - 10^{-5}$ eV) and, sometimes, X-ray bands ($\sim 10^{2.5} - 10^4$ eV).
Consider first RC. 
For electrons emitting in this band, 
the long loss times in any reasonable magnetic field
essentially render our mechanism irrelevant; even considering CGM or cluster scales of 100 kpc, both extremely low scattering rates and uncomfortably large magnetic fields would be required to produce an offset.
It would, moreover, be difficult or impossible to conclusively connect a CR acceleration site to an offset RC source separated by this sort of distance.
Contrastingly, the X-ray band presents a much more favourable case for synchrotron emission:
For emitted photon energies in the $\sim $1-10 keV range and for strong by not implausible magnetic fields in the 5-30 $\mu$G range, \autoref{f:synch} shows that
pitch angle scattering rates close to those inferred for the MW CR population produce offsets in the $\mathcal{O}(10)$ pc range.
Such physical scales are resolvable in the Milky Way by X-ray telescopes by orders of magnitude; indeed, one encounters the opposite problem that they will tend to correspond to angular scales beyond the field-of-view of pointed instruments, so an observing campaign with multiple pointings would be required to bridge between CR source and the displaced emission for MW-plane-typical scattering rates.
On the other hand, we have already mentioned examples of pulsars with bow shocks and magnetotails where displaced X-ray emission seems to have been already identified; this phenomenology tends to require enhanced scattering rates, which is qualitatively reasonable close to energetic injectors. 

Finally, consider the two IC cases illustrated in \autoref{f:CMBIC} and \autoref{f:optIC}.
These show that the natural place to seek displaced $\gamma$-ray sources is in the VHE or above, $\gtrsim$ 10 TeV.
At such energies, pitch angle scattering rates are not too different from those 
inferred for the general Milky Way CR population will produce 
$\gamma$-ray
sources displaced by $\sim 10$ pc from their injection sites for strong, but again not unprecedented, magnetic field amplitudes in the range 10-100 $\mu$G.
With sensitivity into the UHE range -- as conferred by LHAASO or HAWC -- we can even get an observable offset at Milky Way scattering rates for magnetic fields in the $\mu$G range.
Going in the opposite direction into the high-energy regime ($\sim$ GeV regime), the magnetic field amplitudes required to produce an offset for Milky Way-characteristic scattering rates are implausibly high, particularly given the relatively modest angular resolutions of $\gamma$-ray detectors. This is consistent with our conclusions from \autoref{ssec:why_no_GeV} regarding the absence of displaced sources seen by {\it Fermi}-LAT.

Thus the broad conclusion we draw from this analysis is that it is not an accident that the only cases of displaced emission detected so far consist of sources emitting in the X-rays via synchrotron and in TeV $\gamma$-rays via IC. These are the only two mechanisms and energy regimes where, given plausible displacement lengths and CR scattering rates, measurable displacement is likely to occur.

{Finally, we note that all of the above discussion has considered magnetic environments typical of or slightly enhanced with respect to those found in the plane of the Milky Way. There are, however, environments with much stronger, $\sim$mG, magnetic fields, for example starburst galaxies \citep[e.g,.][]{Lacki13a, Krumholz20a} and in the non-thermal filaments seen near the Milky Way centre \citep[e.g.,][]{Morris06a}. For such enhanced field strengths, the corresponding CR energy at which displacement is possible is greatly reduced, and it potentially becomes possible to see displaced inverse Compton off the CMB at $\sim 10$ GeV energies (\autoref{f:CMBIC}) or displaced synchrotron in radio (\autoref{f:synch}). The challenge of course is that these environments are much more distant, and thus correspondingly dimmer and less well-resolved, but they nevertheless present the best prospects for detection of displaced emission at lower energies.}

\subsection{Comparison to alternative mechanisms for displaced sources}

It is interesting to compare the mechanism that we explore here to {two other recent proposed interpretations for displaced $\gamma$-ray sources. The first of these, from \citet{Bao2025}, is that displaced sources appear when} magnetic field lines are bent toward or away from the observer, and electrons propagating along such curved field lines can accumulate along the line of sight, producing projection-enhanced emission that appears spatially offset from the true parent accelerator, thereby generating a ``miraging'' effect. While both this mechanism and ours rely on the presence of strong magnetic fields, the underlying physical interpretations differ in several important respects. Our work models CR transport using a diffusion framework, which captures the collective behavior of electrons undergoing pitch-angle scattering and emphasizes the role of anisotropic particle injection. In contrast, the alternative approach follows the trajectories of individual electrons propagating through synthetic three-dimensional turbulent magnetic fields, with miraging arising primarily from the geometric alignment of magnetic flux tubes with the observer’s line of sight. 

The \citeauthor{Bao2025} scenario also has a weaker dependence on CR energy and is particularly sensitive to the detailed structure of the magnetic field itself. In this alternative scenario, one generically expects the presence of a $\gamma$-ray bridge -- and an even more prominent X-ray bridge -- connecting the parent source to the apparent emission region, as the enhanced brightness arises from projection along magnetic flux tubes aligned with the line of sight. Such connecting structures are expected to be fainter in our framework. Consequently, deeper observations with increased exposure time and improved instrumental sensitivity should provide a clear means of distinguishing between these two mechanisms. Perhaps more significantly, it is unclear why the \citeauthor{Bao2025} mechanism would not operate for GeV emission just as well as for TeV emission. Thus, one might naturally expect that, if miraging rather than our proposed mechanism is the dominant explanation for displaced TeV sources, some of those sources should also show displacement at GeV energies. Careful analysis of \textit{Fermi}-LAT data in the vicinity of these sources thus provides another possible approach to distinguishing between their proposed mechanism and ours.
Additionally, X-ray polarization offers a potential way to distinguish between the two scenarios. In their model, the brightest emission occurs where the magnetic field is nearly aligned with the line of sight, resulting in a small plane-of-the-sky field component and therefore weak linear polarization. By contrast, in our scenario, the magnetic field in general need not be aligned with the line of sight, so the X-ray emission can be strongly linearly polarized. 

{The second recent proposal is from \citet{Sun26a}, who seek to explain a number of cases of displaced emission involving Galactic pulsars, including PSR J1740+1000.  
In their proposed mechanism, CRe are injected with a quasi-isotropic distribution into a straight but expanding magnetic flux tube. However, as the particles propagate along the tube, conservation of magnetic moment in the declining magnetic field amplitude of the expanding tube requires that the distribution becomes concentrated within a progressively narrower pitch angle around $\mu \sim 1$.
For a geometry where the observer is within this cone's opening angle, an IC source will appear to progressively brighten as more and more of the particles propagating down the tube, relatively speaking, start to beam at the observer. 
On the other hand, synchrotron emission along the tube can actually dim for the same geometry, given the decline in the magnetic field.
In distinction to our work, \citet{Sun26a} 
perform their morphological modelling in the limit of zero losses, which is appropriate given that the transport times are significantly shorter than radiative loss times for their preferred parameters and the systems and scales they consider.
In any case, we note that, as for the model of \citet{Bao2025},
the mechanism explored by \citet{Sun26a} predicts, in principle, an IC source morphology that is invariant across the $\gamma$-ray band, and thus there is no obvious reason why displacement should not be as common at GeV energies as at TeV ones. Additionally, for both mechanisms, detectable emission requires the observer’s line of sight to align with a narrow magnetic flux tube, raising questions about the likelihood of such alignment for each observed displaced source.
}

%IVA. General conclusions about when we are expect displacement. Here the argument is basically that the overlapping requirements that (1) the peak be noticeably brighter than emission right on top of the source (so that displacement is detectable), (2) we have a relatively narrow peak (i.e., zeta_max / sigma_zeta is not too small), and (3) the emission be bright enough to be detectable (i.e., f / f_iso is not too small) require that we be observing at mu ~= -0.25 - 0.75, and q ~ 0.1 to a few tenths.
%IVB. When displacement is seen, the fact that the location of the maximum doesn’t depend on much means you can straightforwardly read off the pitch angle scattering rate from the displacement distance. Implications of this for other displaced sources.
%IVB. Why displacement is a phenomenon mostly seen at TeV energies — this is basically an argument that, since you need to have q ~ 0.1, you need to be at energies such that the loss time at reasonable magnetic field strength is comparable to the pitch angle scattering time, but in turn the fact that you can’t measure displacements that are too small (<~ 1 pc) or too large (~100 pc) due to observational constraints means that you need to be operating in a particular CR energy range.
\section{Conclusions}
\label{S:conclusion}

Recent observations have led to the discovery of a class of TeV $\gamma$-ray and X-ray sources where non-thermal emission is visibly displaced from the site of the particle accelerator responsible for powering it. These sources represent a new window into the physics of non-thermal particle transport, since the most plausible explanation for them to date is that they are produced when leptons are injected onto a magnetic field with an anisotropic pitch angle distribution and a correspondingly narrow cone of inverse Compton and synchrotron emission; the particles then travel a finite distance before scattering processes broaden their pitch angle distribution and emission cone enough for the Earth to fall within it. In this paper, we have explored the implications of this picture for the  physical properties of the sources and their utility as a tool for tracing the physics of cosmic ray (CR) transport. Below, we list our three major conclusions.
\begin{itemize}
    \item Our first major conclusion is that visible displaced emission requires a fairly specific set of circumstances. To produce a visible displacement, the maximum of CRe spatial distribution at a particular combination of pitch angle and particle momentum must be spatially offset from the acceleration site by an amount that is not much smaller than the width of the distribution, the CRe density at the peak of the distribution must be significantly higher than that at the acceleration site, and the overall amplitude of the CRe distribution at the position of the maximum cannot be so small as to make their radiation emission unobservable (See \autoref{SS:metrics}, \autoref{sssec:criteria}). All three of these conditions are satisfied only for relatively high-momentum leptons for which the synchrotron or inverse Compton loss time becomes comparable to or smaller than the pitch angle isotropistion time, and for sources where the leptons are injected with a range of pitch angles $\lesssim 45^\circ$ and where the magnetic field is oriented relatively side-on, at angles from $\approx 40^\circ - 110^\circ$ relative to the observer. The spectral shape of the injected leptons has relatively little effect (See \autoref{SS:parameter},\autoref{f:color_map_2}, \autoref{f:par}, \autoref{f:par_2}). 

\item Second, in situations where the conditions for detectable displaced emission are met, the dimensionless displacement distance -- defined as the distance normalised to $c/K_\mu$, where $K_\mu$ is the pitch angle scattering rate -- is always $\approx 0.1$ within a factor of two (See \autoref{f:par_zeta_max}). This means that a detection of displaced emission for a source at known distance can immediately be translated into a rough estimate for the pitch angle scattering rate and the corresponding spatial diffusion coefficient (See \autoref{ss:sc_rate}, \autoref{eq:Kmu_approx}, \ref{eq:Kx_approx}). This provides a new technique to make direct measurements of the pitch angle scattering rate based on first-principles physics, with very few theoretical or modelling uncertainties.

\item Third, the condition that visible displacements happen only when the CR loss time is comparable to the pitch angle isotropisation time naturally explains why displacement can occur for X-ray synchrotron and TeV inverse Compton emission, but for other emission mechanisms or at other energies. For plausible interstellar magnetic field strengths in the Galaxy, the loss times of GeV leptons are long enough that particles either isotropise long before they suffer any significant losses -- in which case we do not expect a bright spot of displaced emission -- or take so long to isotropise that, by the time they do, their emission is displaced so far from the accelerator that it becomes impossible to associate the two. Only at $\gtrsim 10$ TeV energies are loss times small enough to allow detectable displaced emission (See \autoref{ss:otherwavelength}, \autoref{f:optIC}, \autoref{f:CMBIC}). Similarly, loss times for hadrons at plausible interstellar densities are so long that displaced sources would likely be hundreds of pc from their parent accelerators (See \autoref{f:hadron}).

\end{itemize}
This last point explains why displaced sources have not previously been discovered via GeV $\gamma$-ray or radio synchrotron emission, and discoveries are only taking off now, as TeV detector sensitivity undergoes a dramatic increase. It also suggests that this source class represents a significant future opportunity, since the upcoming Cherenkov Telescope Array promises to revolutionise our view of the TeV sky. The theoretical models we have presented in this paper will then provide a crucial tool for the interpretation of new displaced emission sources that CTA uncovers.

\section*{Acknowledgements}

The authors acknowledge support from the Australian Research Council through its \textit{Discovery Projects} scheme, awards DP230101055 and DP260100433. This research was supported by the Australian Government's National Collaborative Research Infrastructure Strategy (NCRIS), with access to computational resources provided by the National Computational Infrastructure through the National Computational Merit
Allocation Scheme, award jh2. MR acknowledges support from the CCAPP fellowship at The Ohio State University and ACCESS allocations (PHY240003). MR acknowledges the Aspen Center for Physics and Simons Foundation, as part of this work was performed there, which is supported by a grant from the Simons Foundation (1161654, Troyer). The analysis in this work was run at facilities supported by the Bridges, University of Pittsburgh, and Anvil, Purdue University. MR thanks Chris Hirata for the useful discussions.

%%%%%%%%%%%%%%%%%%%%%%%%%%%%%%%%%%%%%%%%%%%%%%%%%%
\section*{Data Availability}
The data supporting the plots within this article are available on reasonable request to the corresponding author. The \textsc{criptic} code used for the simulations in this work is available at \href{https://bitbucket.org/krumholz/criptic}{https://bitbucket.org/krumholz/criptic}.



%\begin{thebibliography}{99}
%\end{thebibliography}

%%%%%%%%%%%%%%%%%%%%%%%%%%%%%%%%%%%%%%%%%%%%%%%%%%

%%%%%%%%%%%%%%%%% APPENDICES %%%%%%%%%%%%%%%%%%%%%
\newpage
\appendix
\section{Derivation of the CRe distribution function for isotropic injection}
\label{A:iso_derive}

Our goal in this appendix is to obtain the steady-state, position-integrated solution to the dimensionless Fokker-Planck equation, \autoref{eq:fpe_nondim}, for a case where the distribution function is always isotropic in pitch angle; this corresponds to adopting the limit $K_\mu\to \infty$.

Since we are searching for steady, position-integrated solutions, we first adopt $\partial f/\partial \tau = 0$, and we integrate both sides of \autoref{eq:fpe_nondim} with respect to $\zeta$ from $\zeta=0$ to $\infty$. This gives
\begin{eqnarray}
    0 & = & \mu f(0,\mu,q) + \frac{\partial}{\partial \mu} \left[\frac{3}{2}\left(1-\mu^2\right) \frac{\partial F}{\partial \mu}\right] +
    \frac{\partial}{\partial q} \left[\left(1 - \mu^2\right) q^2 F \right]
    \nonumber \\
    & & \quad
    {} + \dot{\mathcal{N}} \frac{dn}{dq} \Theta(\mu-\mu_\mathrm{inj}),
\end{eqnarray}
where we have defined $F \equiv \int_0^\infty f \, d\zeta$. If injection is isotropic, then we have $\mu_\mathrm{inj} = -1$ and we can drop the final step function term, and if we further assume that $f$ (and thus $F$) are independent of $\mu$, then we can average both sides over $\mu$ to obtain
\begin{equation}
    0 = \frac{d}{dq} \left(q^2 F_\mathrm{iso}\right) + \dot{\mathcal{N}} \frac{dn}{dq},
\end{equation}
where we define $F_\mathrm{iso} \equiv (1/2) \int_{-1}^1 F \, d\mu$, i.e., $F_\mathrm{iso}$ is the pitch angle-averaged, spatially-integrated distribution function. If we now integrate both sides from $q$ to infinity, and assume that $F_\mathrm{iso}\to 0$ as $q\to\infty$, we obtain
\begin{equation}
    q^2 F_\mathrm{iso}(q) = \dot{\mathcal{N}} \int_q^\infty \frac{dn}{dq'} \, dq'.
\end{equation}
Evaluating the integral, recalling that $dn/dq$ is normalised such that $\int_0^\infty (dn/dq)\, dq = 1$, we arrive at the final result,
\begin{equation}
    F_\mathrm{iso}(q) = \frac{\dot{\mathcal{N}}}{q^2} \frac{\Gamma(1+k_p, q/q_\mathrm{cut})}{\Gamma(1+k_p, q_\mathrm{min}/q_\mathrm{cut})},
\end{equation}
where $\Gamma(a,z)$ is the upper incomplete $\Gamma$-function.

\bibliography{main}
\bibliographystyle{aasjournal}
\end{document}